\documentclass[fleqn,10pt]{wlscirep}
\usepackage[utf8]{inputenc}
\usepackage[T1]{fontenc}
\usepackage{multirow}
\usepackage{soul}
\usepackage{cleveref}
\usepackage{algorithm2e}
\usepackage{hhline}
\usepackage{float} 
\usepackage{url}
\usepackage{todonotes}
\usepackage{subcaption}

\newcommand{\June}{{\rm J\scalebox{0.8}{UNE}}\xspace}
\newcommand{\JUNE}{\June}
\newcommand{\JUNECOX}{{\rm J\scalebox{0.8}{UNE}-C\scalebox{0.8}{OX}}\xspace}
\newcommand{\JuneCox}{\JUNECOX}
\newcommand{\JUNEUK}{{\rm J\scalebox{0.8}{UNE}-U\scalebox{0.8}{K}}\xspace}
\newcommand{\JuneUk}{\JUNEUK}

\crefname{table}{Table}{Tables}
\crefname{appendix_table}{Table}{Tables}
\crefname{Table}{Table}{Tables}
\crefname{figure}{Figure}{Figures}
\crefname{Figure}{Figure}{Figures}
\crefname{fig}{Figure}{Figures}
\crefname{Fig}{Figure}{Figures}
\crefname{app}{Appendix}{Appendices}
\crefname{appendix}{Appendix}{Appendices}
\crefname{Appendix}{Appendix}{Appendices}
\crefname{eq}{Equation}{Equations}
\crefname{equation}{Equation}{Equations}
\crefname{Equation}{Equation}{Equations}
\crefname{section}{Section}{Sections}
\crefname{section}{Section}{Sections}

\title{A Mixed-Method Approach to Determining Contact Matrices in the Cox's Bazar Refugee Settlement}

\author[1,2,*]{Joseph Walker}
\author[1,3,**]{Joseph Aylett-Bullock}
\author[1,4]{Difu Shi}
\author[5]{Allen Gidraf Kahindo Maina}
\author[6]{Egmond Samir Evers}
\author[7]{Sandra Harlass}
\author[1,2,***]{Frank Krauss}
\affil[1]{Institute for Data Science, Durham University, Durham, UK}
\affil[2]{Institute for Particle Physics Phenomenology, Durham University, Durham, UK}
\affil[3]{United Nations Global Pulse, New York, USA}
\affil[4]{Institute for Computational Cosmology, Durham University, Durham, UK}
\affil[5]{UNHCR Public Health Unit, Cox’s Bazar, Bangladesh}
\affil[6]{WHO Headquarters, Geneva, Switzerland}
\affil[7]{UNHCR Public Health Unit, Geneva, Switzerland}

\affil[*]{j.j.walker@durham.ac.uk}
\affil[**]{joseph@unglobalpulse.org}
\affil[***]{frank.krauss@durham.ac.uk}

\begin{abstract}
Contact matrices are an important ingredient in age-structured epidemic models to inform the simulated spread of the disease between sub-groups of the population. These matrices are generally derived using resource-intensive diary-based surveys and few exist in the Global South or tailored to vulnerable populations. 
In particular, no contact matrices exist for refugee settlements - locations under-served by epidemic models in general. 
In this paper we present a novel, mixed-method approach, for deriving contact matrices in populations which combines a lightweight, rapidly deployable, survey with an agent-based model of the population informed by census and behavioural data. We use this method to derive the first set of contact matrices for the Cox's Bazar refugee settlement in Bangladesh.  
The matrices from the refugee settlement show strong banding effects due to different age cut-offs in attendance at certain venues, such as distribution centres and religious sites, as well as the important contribution of the demographic profile of the settlement which was encoded in the model. These can have significant implications to the modelled disease dynamics.
To validate our approach, we also apply our method to the population of the UK and compare our derived matrices against well-known contact matrices previously collected using traditional approaches.
Overall, our findings demonstrate that our mixed-method approach can address some of the challenges of both the traditional and previously proposed agent-based approaches to deriving contact matrices, and has the potential to be rolled-out in other resource-constrained environments. 
This work therefore contributes to a broader aim of developing new methods and mechanisms of data collection for modelling disease spread in refugee and IDP settlements and better serving these vulnerable communities.
\end{abstract}

\begin{document}


\flushbottom
\maketitle
\section*{Introduction}
Epidemics such as COVID-19 have led to devastating consequences for afflicted individuals and their societies. 
Understanding how such infectious diseases spread, anticipating future trajectories for transmission, and gathering evidence to inform decision-making efforts to prevent, mitigate and respond to epidemics is therefore of vital importance. 
Mathematical and computational models to simulate disease spread are regularly used to support these efforts.
Contact matrices are key to understanding social mixing patterns in populations, and a vital input to epidemiological models~\cite{fine_measles_1982, anderson_age-related_1985}. 
Despite renewed efforts to develop such models, additional work must be done to ensure they are available to all~\cite{aylett-bullock_epidemiological_2022}.

In this paper we present a new method for determining contact patterns based on combining the information gained from increasingly sophisticated models of disease spread, with that from lightweight surveys which can be rapidly rolled out to populations of interest. 
We attempt to provide information on contact patterns without requiring the traditional, costly methods of contact data collection.
Specifically, we will focus on the use case of the Cox's Bazar refugee settlement in Bangladesh. 
Epidemics in refugee and internally displaced person (IDP) settlements are commonplace and tend to spread rapidly~\cite{altare_infectious_2019}, and only very few models have been designed to simulate outbreaks in these unique environments and to inform public health decision-making~\cite{aylett-bullock_epidemiological_2022}. 
Given the application domain, we believe this is not just an important area in which to contribute knowledge about disease spread patterns, but also a challenging test case which demonstrates the strengths of our methodology.

Throughout this work we will use the \JuneCox model~\cite{aylett-bullock_operational_2021}, an agent-based model built on the \JUNE framework~\cite{aylett-bullock_june_2021}.
The model constructs a virtual population at the level of individual residents within a digital twin of the Cox's Bazar settlement.
Interactions are simulated between the agents -- the virtual residents -- in a number of "venues" or "locations" that include: shelters; food distribution centres; market places; and learning centres.
We use the information from the lightweight survey to guide these interaction patterns based on the demographics of the agents attending the venues contemporaneously.

The contact matrices encode information on the number and duration of contacts between people of one age group and another, and are usually specific to certain venues or locations in which people interact. 
There are various types of matrices which can be used both separately and combined, including (i) one-directional, contact matrices NCM~\cite{mossong_social_2008} which count the (normalised) number of contacts a person in category $i$ has with a person in category $j$, (ii) bi-directional reciprocal matrices NCM$_\textbf{R}$~\cite{mossong_social_2008} which also add the number of contacts people in category $j$ have with persons in $i$, and (iii) venue contact matrices NCM$_\textbf{V}$~\cite{fumanelli_inferring_2012, del_valle_mixing_2007} which assume that every person at venue $L$ has contact with everybody else present.
In this article, we will discuss an approach to estimating all three types of matrices.

Traditionally, contact matrices are derived using large scale surveys in which participants record the number of contacts they have in different locations and the ages of the people they came into contact with. 
Additional metadata is sometimes collected, such as the intensity of the contact (\textit{e.g.}\ physical or non-physical) and the duration of each individual contact. 
Surveys of these types have predominantly been run in the Global North, with comparatively few serving countries in which many particularly vulnerable communities reside~\cite{hoang_systematic_2019}. 
Indeed, to date and to our knowledge only one work has published contact matrices for an IDP settlement~\cite{van_zandvoort_social_2021}, and no such work exists on contact matrices in refugee settlements. 
While such traditional methods of collecting contact data may be considered the gold standard, they are extremely resource consuming to collect, and therefore cannot be run easily during an ongoing outbreak.
As an alternative to these expensive direct means of contact data collection, several other methods have sought a more indirect approach. 
Using the information from existing contact surveys conducted in 8 European countries \cite{mossong_social_2008}, and knowledge of the underlying demographic structures in these populations, Prem \textit{et al.}~\cite{prem_projecting_2017} used a Bayesian hierarchical model to project these matrices onto those of 144 of countries given similar demographic data and underlying similarities between each of these countries and the original 8 selected in the direct data collection.
This has recently been expanded to 177 countries~\cite{prem_projecting_2021}.

Similarly, census/demographic data have also been used to construct synthetic populations which are then used to estimate contact matrices. 
Fumanelli \textit{et al.}~\cite{fumanelli_inferring_2012} use such data from 26 European countries to construct representative synthetic household, school, workplace and `general community' environments and then assume that each individual in each setting has a single contact with every other member.
This has been extended to 35 countries, while also incorporating finer-grained data to develop more representative virtual populations~\cite{mistry_inferring_2021}. 
The same approach is used by Xia \textit{et al.}~\cite{xia_identifying_2013} for the setting of Hong Kong. 
While such approaches are beneficial as they do not require the expensive collection of long-term contact survey data, they are limited by the assumption that different venues contain static populations and that within venue mixing is homogeneous. 

By combining demographic data with data sources such as time use surveys~\cite{zagheni_using_2008, iozzi_little_2010} or transportation surveys~\cite{del_valle_mixing_2007}, stochastic approaches --- \textit{e.g.}\ agent-based models --- have been developed to capture a broader variety of mixing patterns in populations. 
These approaches expand on those described above by exploring many permutations of possible within venue mixing patterns.
Despite this, these methods still present similar limitations as those described above. 
Namely, in the absence of any prior information on interaction patterns, it is largely assumed that each agent contacts every other agent in those venues. 
As a partial remedy to this challenge, disease data is commonly used to fit integer multipliers to these matrices.
While this is generally a necessity to be able to forecast disease spread even when using directly collected contact data~\cite{Vernon2022.02.21.22271249}, due to differences between disease transmission routes this may not resolve the errors at the matrix element level.
Indeed, the output of this process does not provide an understanding of the base level of contacts, but rather a set of contact matrices for each disease. 
This limits the usefulness and generalisability of such matrices in comparison to corresponding matrices from directly collected data.

In this paper, we seek to contribute at two levels: 
i) We develop a methodology which addresses the challenges above by taking a mixed-method approach to deriving contact matrices. 
It combines techniques of extracting contact matrices from sophisticated agent-based models, with information derived from a lightweight survey designed to inform and validate the model-derived matrices, while being significantly less expensive to run than the traditional large-scale contact surveys.  
ii) We use this new approach to present, to the best of our knowledge, the first contact matrices for a refugee settlement. 
Because of their use in different types of models the matrices need different normalization, either to the full population, as in the case of location-unspecific simple compartment models of the SEIR type, or to the part of the population actually visiting a venue.
We will therefore present results for all three types of contact matrices, for a variety of locations, either normalised to the overall population "P" type contact matrices (PNCM, PNCM$_\textbf{R}$, and PNCM$_\textbf{V}$) or to the actual users of a location "U" type matrices (UNCM, UNCM$_\textbf{R}$, and UNCM$_\textbf{V}$).

This work also therefore contributes to the global call to action laid out in prior work, which aims, among others, to develop new methods and mechanisms of data collection for modelling disease spread in refugee and IDP settlements~\cite{aylett-bullock_epidemiological_2022}.
\section*{Methods}

The goal of our method is to construct location-dependent social contact matrices with a high level of granularity without resorting to detailed contact surveys.
We achieve this by fitting the (virtual) contact matrices of an individual-based model constructed from higher-resolution demographic data of the population to the real-world results from lightweight surveys with a much lower resolution.
The resolution and accuracy implicit to the model allows us not only to infer the highly-granular contact matrices, but also allows us to give a first estimate of the associated uncertainties.
In the following we further detail this procedure and exemplify it with the construction of social contact matrices for the residents of Cox's Bazar refugee settlement.

\subsection*{The Survey}\label{sec:theSurvey}
 
The level of detail accessed by surveys in refugee camp settings is often heavily constrained by resource considerations (timing, number of enumerators, need for rapid results etc.), and the highly aggregate contact survey we ran in the Cox's Bazar refugee settlement between October-November 2020 is no exception. 
During this period, the settlement was continuing to experience cases of COVID-19~\cite{who_ewars_20_2022}.
However, reported case numbers were low, and the settlement activity had largely returned to pre-pandemic levels, with the exception that learning centres (schools) remained closed and masks were still being worn~\cite{gbd_2020_2, cbp_interviews}. 
The following demonstrates the ability to rapidly run a survey during a public health emergency, in a resource-light way, while producing representative results of the contact patterns which can be used in future studies and modelling works. 
Although a more intensive survey - such as a diary-based longitudinal study - would provide more precise and accurate data, the ability to perform such a survey may be limited by the number of researchers available or more practical concerns such a limiting social contacts between members of the community and enumerators during a public health crisis. 

The survey underpinning our study was conducted by experienced enumerators from the UNHCR Community Based Protection (CBP) team, following standard UNHCR practices~\cite{unhcr_data_collection, unhcr_data_protection}. Its objective was to collect information on the number of contacts people of different demographics estimate they have with others in different venues they attend during a typical day.
The survey considered only three categories of residents, defined by their age: children ($< 18$ years), adults ($\geq 18$ and $< 60$ years), and seniors ($\geq 60$ years), and we constrained the set of surveyed locations to those contained in the digital twin, \JuneCox.
Data was collected from 22 camps in the Kutapalong-Balukhali Expansion Site (part of the Cox's Bazar refugee settlement).
In each camp the respondents were two male and two female residents in each of the three age brackets. 
In addition, two persons with disabilities were surveyed in each camp, resulting in a total of $22\times 14 = 308$ respondents. 
Details of the survey can be found in Appendix~\ref{app:survey} and the accompanying metadata to the anonymised results~\cite{microdata_survey}.
The respondents were asked if they attend various venues and, if so, to estimate the number of adults and children they come into contact with there. 
To avoid skewing results through uncharacteristically long or short times at a venue, the respondents were asked how much time they generally spend at those venues at any given visit such that the total contacts can be re-scaled to contacts per hour. 
Since the \JUNE modelling framework normalises the contact matrices to represent the mean rate of contacts per hour, many of the demographic data underpinning \JuneCox do not distinguish adults and seniors and so we combine the data in these two age bins into one "adult" category, thereby arriving at highly aggregate $2\times 2$ total contact contact matrices $t_{ij}$\footnote{We interpret contact matrix $\Delta_{ij}$ such that person $i$ contacts person $j$ and graphically as subgroup on $x$-axis contacts subgroup on $y$-axis.} for the various locations $L$\footnote{To improve the readability of the manuscript we refrain, where possible, from explicitly indexing contact matrices etc.\ with a location index.}.
We use the survey to calculate UNCM$_\textbf{R}$ type matrices for different locations. 
Here we present the methodology to calculate the different versions of the contact matrices:
\begin{enumerate}
    \item One-directional contact matrices~\cite{mossong_social_2008},  NCM, (UNCM and PNCM):
    Following the notation in~\cite{klepac_contacts_2020} the PNCM are denoted as $M$ with elements $m_{ij}$ defined by $
        m_{ij} = t_{ij}/n_j
    $
    with $t_{ij}$ the aggregate total number of contacts of $n_j$ survey respondents in category $j$ reported with people in category $i$.
    
    There is a subtle difference to the UNCM with elements $\mu_{ij}$, where the aggregate number of contacts $t_{ij}$ is normalised to the number of 
    actual users in the venue, $\eta_j$
    $
        \mu_{ij} = t{ij}/\eta_j$.
    To make contact between the PNCM and UNCM, one therefore merely has to re-normalise to the overall number of respondents in category $j$, $
        m_{ij} = t_{ij}/n_j = 
        \mu_{ij} \eta_j/n_j =
        \mu_{ij} a_j$,
    where $a_j$ denotes the attendance rate to the venue in category $j$. This re-normalisation can be performed for any conversion from population normalised "P" to user "U" normalised matrices.
    \item Bi-directional, reciprocal contact matrices~\cite{mossong_social_2008}, NCM$_\textbf{R}$, (UNCM$_\textbf{R}$ and PNCM$_\textbf{R}$): 
    Following, again~\cite{klepac_contacts_2020}, the PNCM$_\textbf{R}$ are denoted by $C$ and their elements are defined as
    \begin{equation}
        \label{Eq:cij_from_BBC}
        c_{ij} = \frac{1}{2}\,
        \left(m_{ij}+m_{ji}\frac{w_i}{w_j}\right) =
        \frac{1}{2w_j}
        \left(t_{ij}\frac{w_j}{n_j} +
        t_{ji}\frac{w_i}{n_i}\right)\,,
    \end{equation}
    where the $w_{i,j}$ are the overall population sizes in categories $i$ and $j$.  
    This motivates the notion of these matrices being normalised to the overall population.
    While using these matrices in compartment models, their application in individual-based models may lead to unwanted results.
    As an example consider the case of contacts between adults and children in school settings, and assuming that this is meant to primarily capture the contact of teachers and pupils.
    Normalising the number of contact to the overall adult population size would obviously lead to a massively reduced average number of contacts compared to a more correct normalization to the number of teachers in the respective age bins.
    We therefore define the user-normalised contact matrices UNCM$_\textbf{R}$ $\Gamma$ with entries
    \begin{equation}
        \label{Eq:cij_from_survey}
        \gamma_{ij} = 
        \frac{1}{2\omega_j}
        \left(t_{ij}\frac{\omega_j}{\eta_j}+t_{ji}\frac{\omega_i}{\eta_i}\right)\,,
    \end{equation}
    where $\omega_{i,j}$ denote the actual users attending the venue, \textit{i.e.}\ $\omega_i = w_ia_i$.
    In fact, since we resolve the random movement of individuals to distinct locations in \JUNE, we use the $\Gamma$ instead of the $C$ that are more relevant for compartment models.
    However, we also present results for the population-normalised PNCM$_\textbf{R}$, which can be obtained by simple rescaling by attendance factors $a_i$ and $a_j$ from the $\Gamma$.
    \item Isotropic venue contact matrices,  NCM$_\textbf{V}$, (UNCM$_\textbf{V}$ and PNCM$_\textbf{V}$): due to the lack of attendance data we cannot directly derive such matrices $v_{ij}$ and $\nu_{ij}$ from the survey. However they can be determined virtually.
\end{enumerate}
Finally, we comment to our treatment of the uncertainties in the survey results. 
Given the small survey sample size, we right-censor the data at the level of the 90th percentile and perform a bootstrap analysis~\cite{Efron_bootstrap_1993} to determine the median number of contacts between subgroups, $\mu_{ij}$.
We assume the uncertainty of this value, $\Delta\mu_{ij}$, to be well estimated by the standard error of the bootstrap distribution.
From $\Delta\mu_{ij}$ it is straightforward to derive the uncertainty, $\Delta\gamma_{ij}$, of the reciprocated matrices, we assume the error in the contacts are dominated by the error from reported number contacts per hour at a venue. We take $\omega_{i,j}$ in Eq.~(\ref{Eq:cij_from_survey}) as an exact quantity from the survey.

\begin{table}[]
    \begin{center}
    \begin{tabular}{|l||l||l|l|}
    \hline
        matrix & symbols & matrix & symbols \\ \hline \hline 
        &&& \\
        CM & t, $t_{ij}$ & & \\ & & & \\
        \hline
        &&&  \\
        UNCM & $\mu$, $\mu_{ij}$ & PNCM & M, $m_{ij}$    \\
        &&&  \\
        UNCM$_\textbf{R}$ & $\Gamma$, $\gamma_{ij}$ & PNCM$_\textbf{R}$ & C, $c_{ij}$   \\ &&& \\
        UNCM$_V$ & $\nu$, $\nu_{ij}$ &  PNCM$_\textbf{V}$ & V, $v_{ij}$\\ &&& \\ \hline  &&& \\
        Population venue & $\eta_{ij}$ & Population world & $\omega_{ij}$ \\ &&& \\
        Population survey venue & $n_{ij}$ &
        Population survey world & $w_{ij}$ \\ &&& \\
        \hline
    \end{tabular}\\[2mm]
    \parbox{0.8\textwidth}{\caption{\label{tab:my_label} The CM are time normalised, the various UNCM are further normalised by population at the venues, while the PNCM are instead normalised by the {\em total} population.}}
    \end{center}
\end{table}

\subsection*{The Model}
For the construction of the digital twin and simulator we use an existing individual-based model, \JuneCox~\cite{aylett-bullock_operational_2021}, specifying the original \JUNE modelling framework~\cite{aylett-bullock_june_2021} to the demographics of the Cox's Bazar refugee settlement. (Note that the original application of the \JUNE framework was to model the spread of COVID-19 in the UK and we will refer to this UK specific specification as \JuneUk.)
Both \JuneUk and \JuneCox use census data to create a virtual population at the individual level, with \JuneCox specifically focusing here on the Kutapalong-Balukhali Expansion Site of Cox's Bazar.
The census data of its population is organised according to a geographical hierarchy; the $\sim$600,000 residents are distributed over the 21 camps ("regions") which make up the Kutapalong-Batukhali Expansion Site (in reality there are 22, however, we combine Camp-20 and the Camp-20 extension together given data availability constrains), these contain between $2$-$7$ UNHCR Admin level-2 blocks ("super areas") comprising $\sim 5000$ people, which in turn are composed of sub-blocks ("areas") with $90$ households on average. 
The geographical distribution of individuals and their households is explicitly incorporated in the model through the geo-locations of the area centres. 
For a more complete description of how we distribute individuals into households see Appendix~\ref{app:DemographicProperties} and the original work describing \JuneCox~\cite{aylett-bullock_operational_2021}.

\begin{figure}[tbp]
\centering 
\includegraphics[width=.9\linewidth]{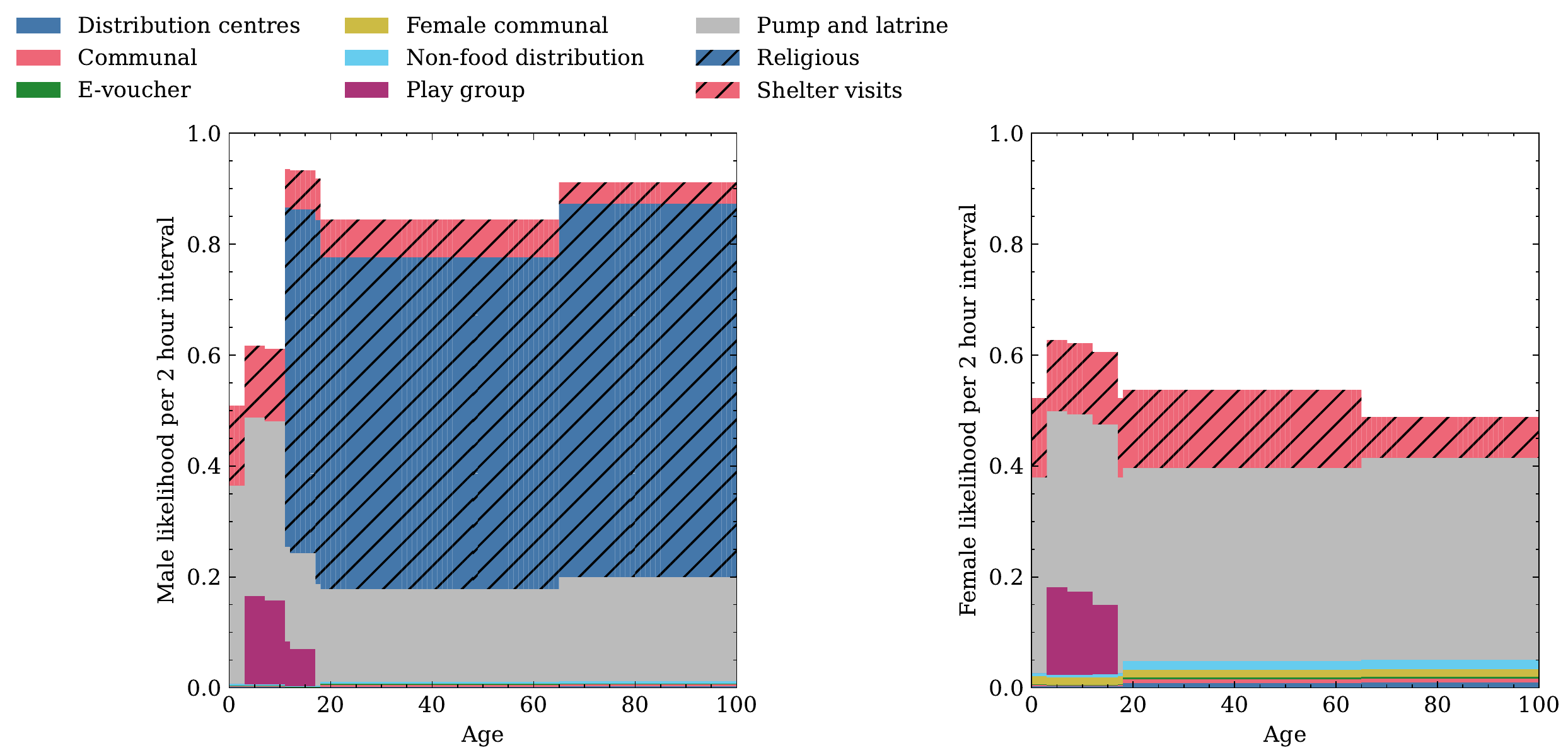}
\parbox{0.8\linewidth}{\caption{\label{fig:AgeProbabilities} The mean likelihood to attend certain venues in any weekday 2 hour timestep interval by age for Left: Men, Right: Women.}}
\end{figure} 
After the individuals are created and clustered into households, \JuneCox constructs different venues in the settlement given their latitude and longitude coordinates: food distribution centres; non-food distribution centres (including LPG distribution centres); e-voucher outlets; community centres; safe spaces for women and girls, religious centres, learning centres, hand pumps and latrines.
To simulate the movement of individuals in the settlement we decompose each calendar day into discrete time-steps in units of single hours. 
\JUNE uses calendar days to distinguish weekday and weekend activity profiles where certain venues will be closed. 
Many individuals have fixed, static, activities, such as the 4 hours at the learning centres for enrolled children and the adults specified as teachers. 
There is also a fixed 14 hours night-time period, during which everyone returns to their shelter. 
However, the remaining time is free and people are distributed dynamically.
Each person not otherwise occupied (\textit{e.g.}\ working, or at a medical facility) is assigned a set of probabilities for undertaking other activities in their free time in the model. 
These probabilities are part of our social interaction model, and depend on the age and sex of the person (Figure~\ref{fig:AgeProbabilities}). 
They are based on previously collected data capturing daily attendance rates and coarse estimates in proportions of adult/child and male/female attendance (see previous work for details on these calculations and associated data sources~\cite{aylett-bullock_operational_2021} and have been further augmented by a series of interviews with CBP officials as detailed in Appendix~\ref{app:Questions}). 

Given $N$ possible activities with associated probabilities per hour given by $\lambda_1 , . . . , \lambda_N$ , for a person with characteristic properties {$p$}, the overall probability, $P$, of an individual being involved with any activity in a given time interval $\Delta t$ is modeled through a Poisson process:
\begin{equation}
\label{Eq:ProbVenue1}
    P = 1 - \textrm{exp}\bigg( - \sum^N_{i=1} \lambda_i({p})\Delta t \bigg)\,.
\end{equation}
If the individual participates in at least one of these activities, the specific activity $i$ is selected
according to:
\begin{equation}
\label{Eq:ProbVenue2}
    P_i = \dfrac{\lambda_i({p})}{\sum^N_{i=1} \lambda_i({p})}\,,
\end{equation}
and the person is moved to the relevant location.
If no activity is selected, the individual will stay in their shelter.
One of the outcomes of this exercise is condensed in Figure~\ref{fig:AgeProbabilities}, which shows the likelihoods that men and women attend the different venues in the model as a function of their age.

It is important to stress that such census and demographic data is by default recorded by UNHCR and other non-governmental organisations (NGOs) operating in refugee and IDP settlements, and it can be further supplemented or clarified by the survey described above or by interviews with settlement staff. 
This implies that it is a relatively straightforward exercise to apply our procedure outlined here to other settlements.

\subsection*{A Mixed-Method Approach}

We have now set the stage to combine the information about the aggregate contact patterns with our highly-detailed model of interactions in a representative virtual population and to interrogate the model and extract detailed, survey informed, matrices.
\JUNE uses stochastic methods to simulate contacts between members of the virtual population which can be used to construct synthetic CMs. 
The random behaviour of the virtual population is encoded in repeatedly sampling the $\gamma_{ij}$ from a Poisson distribution,
$\tilde{\gamma}_{ij} \sim \mathcal{P}(\kappa_{ij})$
with the argument $\kappa_{ij}$ distributed according to a normal distribution,
\begin{equation}
    \kappa_{ij} \sim \mathcal{N}\left(\bar\mu,\,\sigma\right)\;\;\;\mbox{\rm with}\;\;\;
    \mu =\dfrac{T}{\Delta T}\gamma_{ij}
    \;\;\;\mbox{\rm and}\;\;\; \sigma=\dfrac{T}{\Delta T}\Delta \gamma_{ij}\,,
\end{equation}
with the $\gamma_{ij}$ and their uncertainty taken from the survey and re-scaled by the ratio of the typical time people attend a location, $T$, and the size of the emulation time-step in the model, $\Delta T$.
Finally we statistically round the individual instances $\tilde{\gamma}_{ij}$ to integer values.
The resulting emulated set of $\tilde{\gamma}_{ij}$ are normalised such that they represent an individual's contacts per hour.
Averaging generates the $\hat{\gamma}_{ij}$ which can be directly compared with the $\gamma_{ij}$ obtained from the survey.

In the simulation we aim to perform a virtual survey on the virtual population, as close as possible to the conditions in the real-world light-weight surveys.
We sample individual behaviour over 28 virtual days to obtain individual $\tilde{\gamma}_{ij}$'s every time a person attends a venue.
The venues are filled according to the probabilities described above, Eqs.~(\ref{Eq:ProbVenue1}, \ref{Eq:ProbVenue2}) and we "measure" the total raw contacts $\hat{t}_{ij}$(see Algorithm~\ref{alg:CountContacts} in Appendix~\ref{app:algo_virtual_survey}) in the simulation.
To further insure the correct total expected attendance time at the virtual venues compared with the real world, we proportionally close venues to approximate their possible fractional opening times.

This procedure allows us to directly compare resulting matrices $\hat{t}_{ij}$, $\hat\gamma_{ij}$, and $\hat{c}_{ij}$ with their real-world counterparts $t_{ij}$, $\gamma_{ij}$, and $c_{ij}$ above.
Even more, we are not constrained to the creation of virtual $2\times 2$ contact matrices only, but can infer matrices for any sub-classification $i$ and $j$ that our simulation allows -- in the results we present here, the $i$ and $j$ are age brackets of size 1 year.
The final type(s) of contact matrix, PNCM$_V$ and UNCM$_V$, $\hat{v}_{ij}$ and $\hat{\nu}_{ij}$, can also be calculated with a minor modification to the algorithm that counts the averaged total contacts per hour, Algorithm~\ref{alg:CountContacts}. 
Instead of generating a list of people ${p_j}$ at the venue in contact with each person $p_i$, we allow "democratic/isotropic" contacts of all people:
\begin{equation}
    \hat{\nu}_{ij} = \frac{1}{\hat{\eta}_i}\left(\phantom{\frac12}\hat{\eta}_i(\hat{\eta}_j-\delta_{ij})\right)\,.
\end{equation}
For each entry, $ij$, this represents the total contacts the $\hat{\eta}_i$ people with characteristics $i$ at the venue have with the population of the venue in each subgroup.  
The Kronecker-$\delta$ corrects for "self-contacts".
\section*{Results}\label{sec:results}

In this section we present the results of the contact matrices derived from our mixed-method approach. 
We begin by validating our method in the context of the UK where we compare our results against contact patterns directly collected by a traditional survey~\cite{klepac_contacts_2020}. 
Once our method has been validated, we present the matrices for the Cox's Bazar refugee settlement. 
Throughout, we use several key metrics to determine the similarity between any two sets of matrices: 
\begin{enumerate}
\item Normalised Canberra distance, $D_C$~\cite{10.1093/comjnl/9.1.60}:
\begin{equation}
    D_C(C,C^{\prime}) = \dfrac{1}{\textrm{Dim}(C) - Z} \sum_i \sum_j \dfrac{|C_{ij}-C^{\prime}_{ij}|}{|C_{ij}|+|C^{\prime}_{ij}|},
\end{equation}
where $C$ and $C^{\prime}$ represent two contact matrices we wish to compare, Dim denotes the number of elements, $\textrm{Dim}(C_{n \times m}) = n \cdot m$, and $Z$ is the number of non zero elements of the difference $(C_{ij}-C^{\prime}_{ij})$.;
\item $Q$ index as measure of assortativity~\cite{Gupta1989NetworksOS}:
\begin{equation}
    Q = \dfrac{\textrm{Tr}(C / \sum_{ij} C_{ij}) - 1}{\textrm{Dim}(C) - 1};
\end{equation}
\item Dissimilarity index, $I^2_s$~\cite{FarringtonDisassortiveness}:
\begin{equation}
    I^2_s = \dfrac{1}{2} \dfrac{\langle ( S-T )^2 \rangle_{F_c}}{\sigma_{p}^4},
\end{equation}
where $\sigma_p$ is the standard deviation of the ages of the population, and $\langle ( S-T )^2 \rangle_{F_c}$ represents the expectation age difference between contacts $s$ and $t$ of the function $F_c(s,t)$:
\begin{equation}
    F_c(s,t) = \dfrac{f(s) C_{st}^L f(t) }{\sum_s \sum_t f(s) C_{st}^L f(t) \Delta t \Delta s }.
\end{equation}
Here, $\Delta t$ and $\Delta s$ are the age bin sizes from the contact survey. 
\end{enumerate}
The normalised Canberra distance gives an estimation of the similarity between two matrices - approaching 0 when they are more similar and 1 when dissimilar.
The remaining statistics measure the level of assortativity - the level of diagonal dominance and therefore the rate at which similar ages interact compared with dissimilar ages. 
The $Q$ index ranges from 0 - homogeneous, proportionate mixing - to 1 - fully assortative. $I^2_s$ measures the deviation from perfect assortativity with a value of 0 when fully assortative, and 1 for homogeneous interactions. 

\subsection*{UK Validation}
The first step of our virtual survey validation is to compare our results with that of real surveys conducted in far greater granularity. 
\JUNEUK has had extensive tuning for COVID-19 modelling in the UK~\cite{aylett-bullock_june_2021, Vernon2022.02.21.22271249, Cuesta-Lazaro2021.09.07.21263223}.
As a proof-of-concept, we focus on the most complex contact matrix -- that of the household -- and compare the contact matrices produced by the simulation with those from a traditional diary-based survey~\cite{klepac_contacts_2020}. 
The input contact matrix is constructed from a combination of this data, the Office of National Statistics (ONS) census data of UK households~\cite{ONS_UK_Households, ONS_UK_NDependents} and UK population demographics~\cite{ONS_UK_PopCenus}.
Since the UK census for household types distinguishes children (kids, K, <18 years old), young adults such as students or other dependent resident (Y, assumed 18-25 years old in \JUNEUK), adults (A, assumed 26-65 years old), and older adults (O, assumed >65 years old), we aggregated the granular contact matrix derived from the survey into a significantly coarser $4\times 4$ matrix mapping the census categories. 
We also corrected for different household types to better incorporate the details of the venue-specific heterogeneities in their demographic composition.  
For more details on this procedure, see specifically Section 4 and Appendix C of the original description of the \JUNEUK modelling setup \cite{aylett-bullock_june_2021}.
\begin{figure}[h!]
    \begin{center} 
    \includegraphics[width=\linewidth]{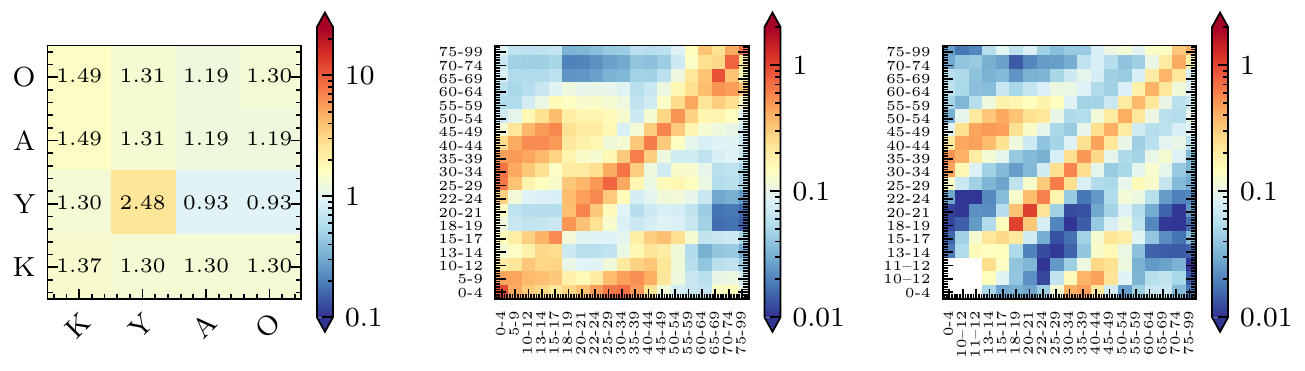}
    \parbox{0.8\linewidth}{
        \caption{\label{fig:CM_UKResult_Household} 
        Left: The derived input interaction matrix, UNCM$_\textbf{R}$ for "Households". 
        Center: The simulated age-\-binned PNCM$_{\textbf{R}}$ matrix with entries $\hat{C}_{ij}$ from \JUNEUK.  
        Right: The BBC Pandemic project~\cite{klepac_contacts_2020} "all home" contact matrix, $C$, with entries $c_{ij}$.}}
    \end{center}
\end{figure} 

The results in Figure~\ref{fig:CM_UKResult_Household} show the $4\times 4$ input matrix derived from the aggregation process described above and a comparison of the output of the PNCM$_R$ $\hat\Gamma$ from the \JUNEUK model virtual contact survey with the results of the matrix $C$ from the traditional survey.
Corresponding results for work place and School settings can be found in Appendix~\ref{app:UK_Val}.
This provides a closure test ensuring that \JUNEUK returns realistic contact matrices from coarse aggregate matrices.  
Clearly, our mixed-method approach is able to reproduce the broad structure of the real-world data - especially capturing the patterns of contacts between children and their parents represented in the off-diagonal structures. 
The original survey did not contain information on the contacts of younger children due to constraints on the data collection methodology; our method is able to fill this gap.
\begin{table*}[h!]
    \begin{center}
    \begin{tabular}{ | c || c | c | c | } 
        \hline
        \parbox{0.1\textwidth}{$\vphantom{\int\limits_{1}^{12}}$} & $Q$ & $I^{2}_{s}$ & $D_{C}$ \\
        \hline \hline
        $\vphantom{\int\limits^1}$BBC Pandemic & $0.14$ & $0.36$ & \\[2mm]
        \hline 
        $\vphantom{\int\limits^1}$\JUNEUK PNCM$_{\textbf{R}}$ & $0.12$  & $0.30$ & $0.32$\\[2mm]
        \hline
    \end{tabular}
    \\[2mm]
    \parbox{0.8\linewidth}{
        \caption{\label{tab:CMStatistics}
        Contact matrix statistics calculated for the results for the a range of contact matrix types from \JUNEUK and the BBC Pandemic project~\cite{klepac_contacts_2020} in Fig.~\ref{fig:CM_UKResult_Household}.
        }
    }
    \end{center}
\end{table*}
To further validate our approach, we compare the $Q$, $I^2_s$ and $D_C$ metrics of the two matrices. 
Table~\ref{tab:CMStatistics} shows that the first two metrics are in close agreement, with the overall Canberra distance being close to 0, thereby confirming the similarity of the matrices. 
Indeed, the difference between the measures of assortativity are comparable or better than those found in similar studies but which do not make use of the guiding input aggregate matrix as we do here~\cite{iozzi_little_2010}.
Given these strong findings, together with the visual and structural similarities of the matrices, we consider our mixed-method approach to be reasonably validated for application to settings in which intensive survey-based approaches to deriving contact patterns are not feasible.
For real-world applications, we note that our methodology is clearly not exactly reproducing the original surveys; however, users will have to decide whether these errors are acceptable in comparison to having little or no knowledge about contact patterns, or making necessary assumptions about these patterns.
It is also worth noting that the virtual agent behaviour of \JUNEUK are much better informed than those in \JUNECOX. This will become clear in the disparity between NCM, NCM$_\textbf{R}$ and  NCM$_\textbf{V}$ type contact matrices. PNCM$_\textbf{V}$ matrices presented in the Figure ~\ref{fig:UK_PNCM_V_syoa} and PNCM$_\textbf{R}$ matrices in Figures~\ref{CM_UKResult_Household}, \ref{fig:CM_UKResult_companySchool} have the same general shape and scaling of features. In the case of \JUNECOX derived matrices NCM, NCM$_\textbf{R}$ and  NCM$_\textbf{V}$ types are less similar.

\subsection*{Contact Matrices in Cox's Bazar Refugee Settlement}
\begin{figure}[h!]
    \begin{center}
    \includegraphics[width=.75\linewidth]{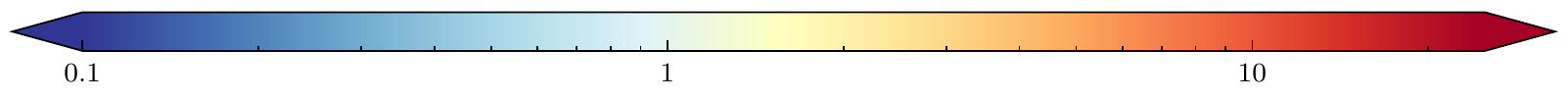} \\[2mm]
    \begin{tabular}{ccccc}
     \includegraphics[width=0.2\linewidth]{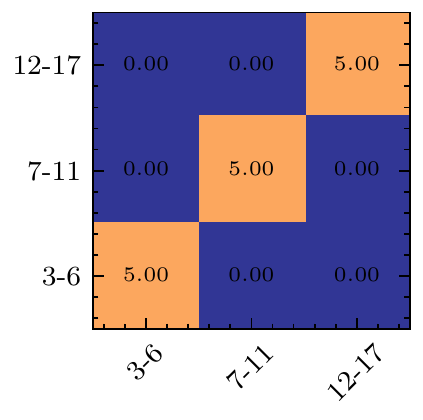} &
     \includegraphics[width=0.2\linewidth]{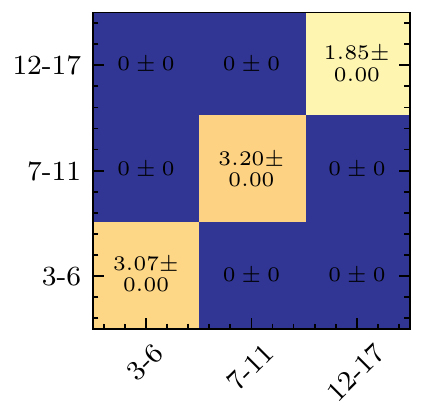} &\hspace*{2mm} &
     \includegraphics[width=0.2\linewidth]{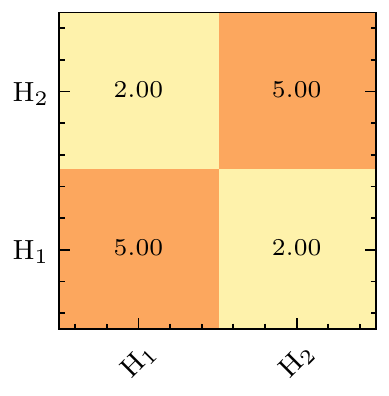} &
     \includegraphics[width=0.2\linewidth]{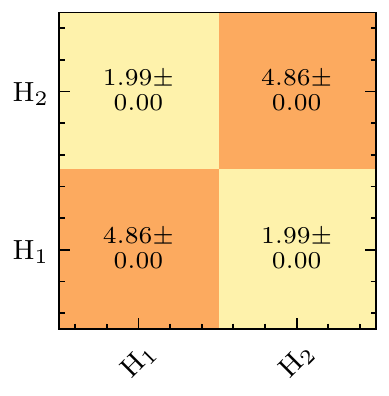}
    \\[-2mm]
     \multicolumn{2}{c}{Play groups: $D_C =  7.8 \times 10^{-3}$} 
    &&
    \multicolumn{2}{c}{Shelters: $D_C =  9.0 \times 10^{-3}$}
    \end{tabular}
     \\[2mm]
    \parbox{0.8\textwidth}{\caption{    \label{fig:SurveyOtherCMresults} 
    Two pairs of UNCM$_\textbf{R}$ for the virtual venues determined prior to the light-weight survey (Left) and the \JUNECOX virtual survey in the same coarse population bins (Right), including the Canberra distance between them.} 
    }
    \end{center}
\end{figure}

\begin{figure}[h!]
\begin{center}
    \includegraphics[width=.75\linewidth]{figures/colourbar_UNCM_Interaction.pdf}
    \begin{tabular}{ccccc}
    \includegraphics[width=0.2\linewidth]{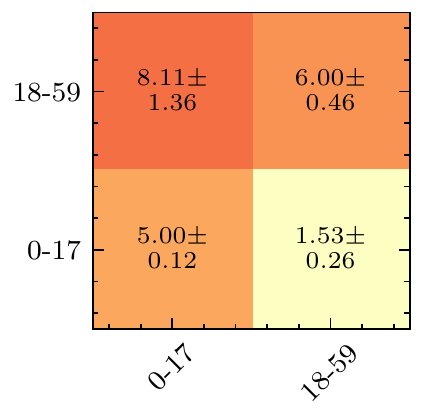} &
    \includegraphics[width=0.2\linewidth]{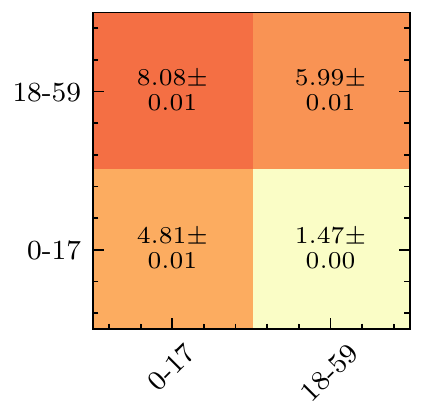} &\hspace*{2mm} &
    \includegraphics[width=0.2\linewidth]{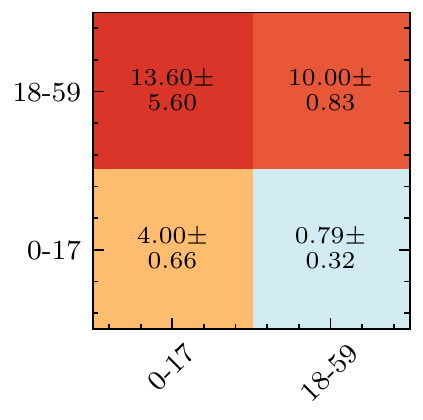} &
    \includegraphics[width=0.2\linewidth]{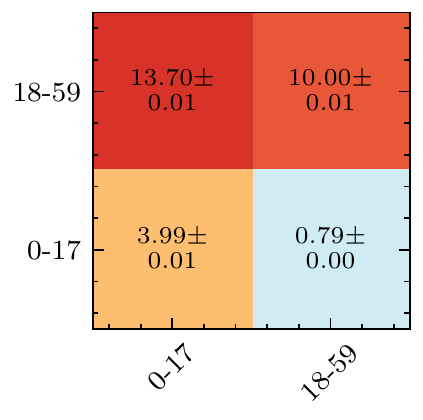} \\[-2mm]
    \multicolumn{2}{c}{Community centres: $D_C =  1.0 \times 10^{-2}$} &&
    \multicolumn{2}{c}{Distribution centres: $D_C =  1.6 \times 10^{-3}$}\\[4mm]
    \includegraphics[width=0.2\linewidth]{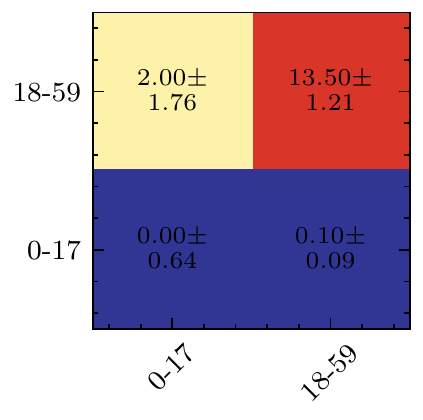} &
    \includegraphics[width=0.2\linewidth]{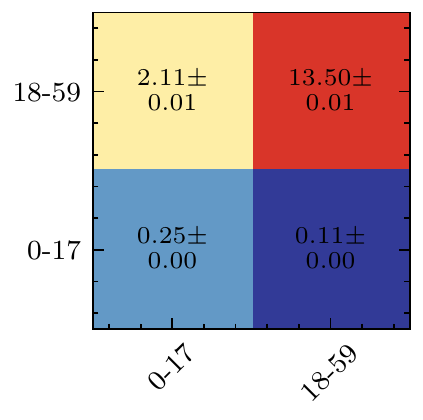} &&
    \includegraphics[width=0.2\linewidth]{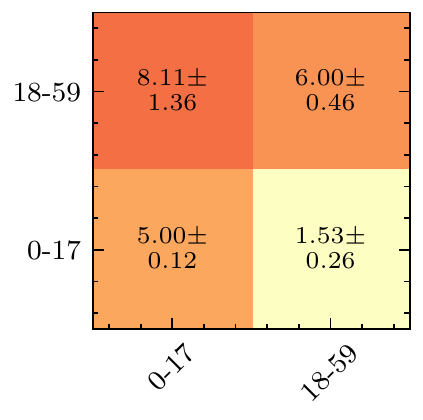} &
    \includegraphics[width=0.2\linewidth]{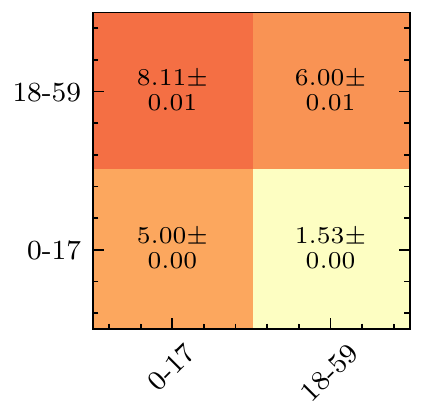}\\[-2mm]
    \multicolumn{2}{c}{e-voucher outlets: $D_C = 0.26$} &&
    \multicolumn{2}{c}{Female friendly spaces: $D_C =  5.6 \times 10^{-4}$} \\[4mm]
    \includegraphics[width=0.2\linewidth]{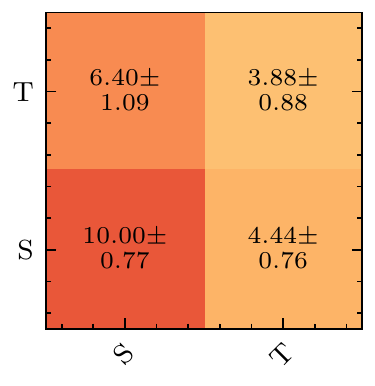} &
    \includegraphics[width=0.2\linewidth]{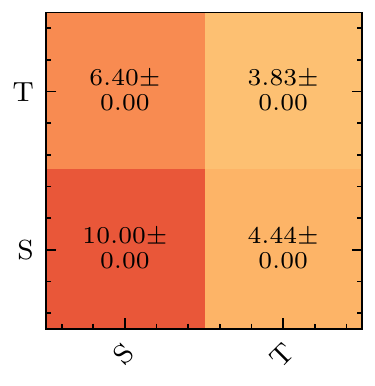} && 
    \includegraphics[width=0.2\linewidth]{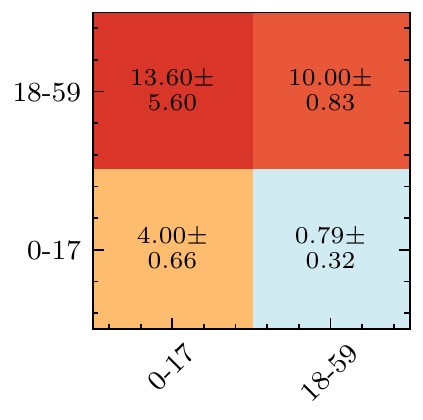} &
    \includegraphics[width=0.2\linewidth]{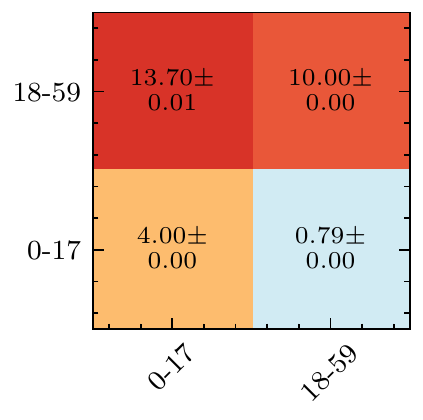} \\[-2mm]
    \multicolumn{2}{c}{Learning centres: $D_C =  1.7 \times 10^{-3}$} &&
    \multicolumn{2}{c}{Non-food distribution centres: $D_C =  7.7 \times 10^{-4}$} \\[4mm]
    \includegraphics[width=0.2\linewidth]{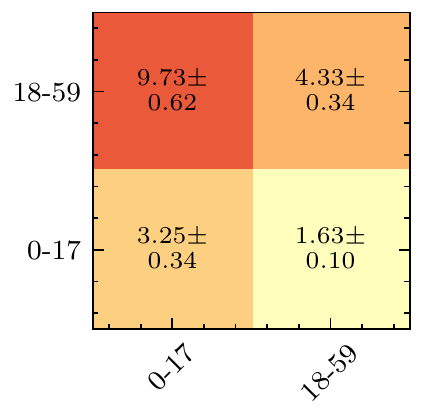} &
    \includegraphics[width=0.2\linewidth]{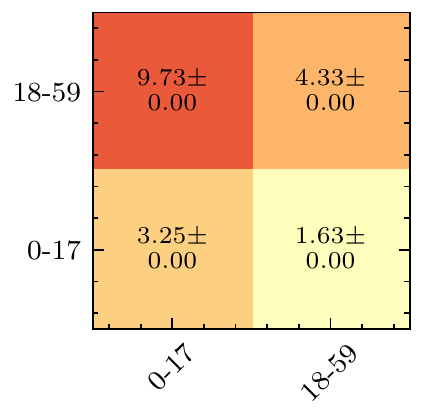} &&
    \includegraphics[width=0.2\linewidth]{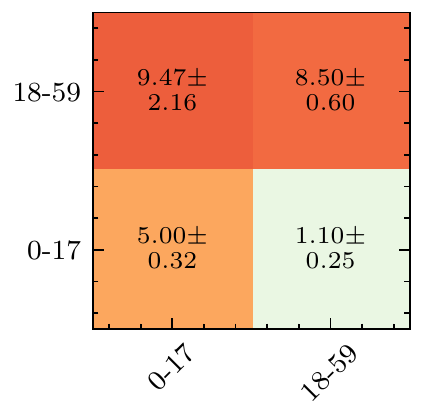} &
    \includegraphics[width=0.2\linewidth]{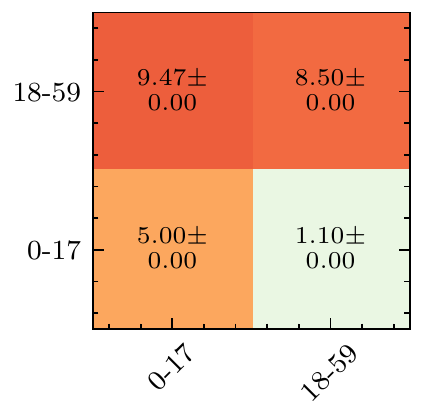} \\[-2mm]
    \multicolumn{2}{c}{Pumps and latrines: $D_C =  5.9 \times 10^{-5}$} &&
    \multicolumn{2}{c}{Religious centres: $D_C =  7.1 \times 10^{-5}$}
    \end{tabular}
    \\[4mm]
    \parbox{0.8\textwidth}{\caption{\label{fig:SurveyCMresults}
    The UNCM$_{\textbf{R}}$ from the contact survey data (Left) and \JUNECOX virtual survey UNCM (Right), with the relative Canberra distances.   
    We set "Community centres" and "Distribution centres" identical to "Female friendly spaces" and "Non-food distribution centres", respectively. } }
\end{center}
\end{figure}

The lightweight survey in the camp was conducted across the following venues: "community centres", "distribution centres", "e-voucher outlets" and "formal education centres".
For the remaining two venues - "play groups" and "shelters" - we assume that everyone generally mixes with everyone else in that location given the assumed small groups of children who play together, as well as the dense shelter environments.
Since certain shelters are shared between multiple families, we differentiate intra- and inter-family mixing with the latter being represented by the diagonal elements of the aggregate matrix (\textit{i.e.}\ setting these to the number of contacts within each of the two families in the shelter, and with the off diagonal elements set to the number of contacts between the families). 
As discussed in previous work~\cite{aylett-bullock_operational_2021}, we set the number of contacts within the families or play groups to the average size of these respective groups assuming homogeneous mixing in these settings.
In the case of the play groups we dis-aggregate the population into three age groups 3-6, 7-11 and 12-17 which mix homogeneously to emulate children typically interacting with children of similar age. 
We report the results for the UNCM$_\textbf{R}$ $\gamma_{ij}$ of the prior information and of the survey in Figs~\ref{fig:SurveyOtherCMresults} and \ref{fig:SurveyCMresults}.
We also perform a closure test by comparing them to the UNCM $\hat{\mu}_{ij}$ results from performing a similar survey in \JUNECOX with the same coarse population categories.
In the two figures we use the shorthand "T" and "S" for teachers and students in the learning centres, and "H$_x$" for household $x$ in a shared shelter.


\begin{figure}[h!]
\centering
\includegraphics[width=.9\linewidth]{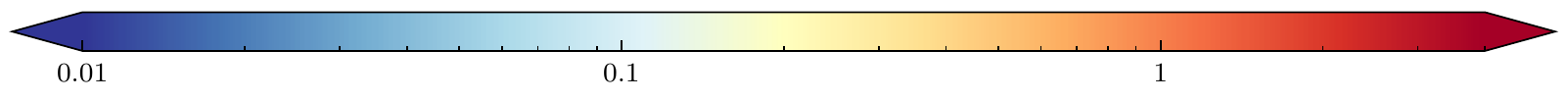}
\begin{tabular}{llllllllllll}
\multicolumn{3}{l}{\includegraphics[width=0.225\linewidth]{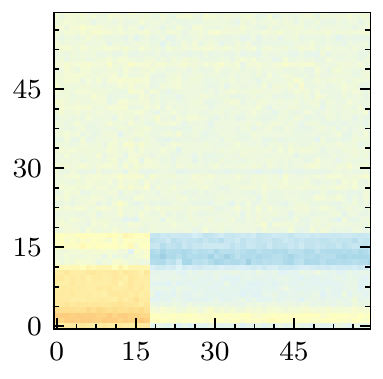}} &
\multicolumn{3}{l}{\includegraphics[width=0.225\linewidth]{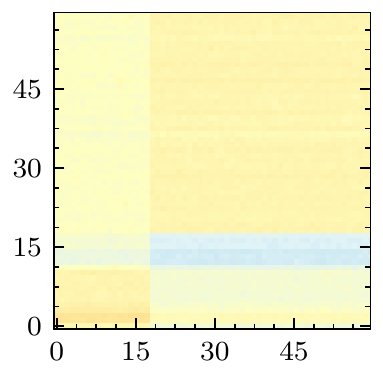}} &
\multicolumn{3}{l}{\includegraphics[width=0.225\linewidth]{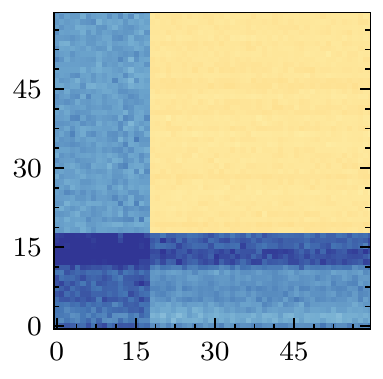}} &
\multicolumn{3}{l}{\includegraphics[width=0.225\linewidth]{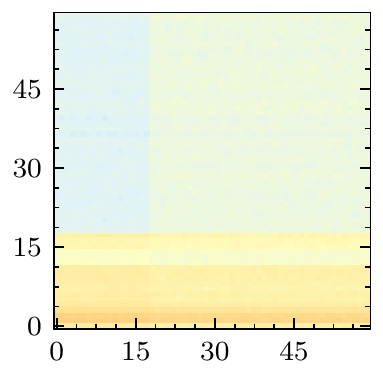}}\\[-1mm]
\hspace*{2mm} & (a) & Community centres: &
\hspace*{2mm} & (b) & Distribution centres:  &
\hspace*{2mm} & (c) & e-voucher outlets:  &
\hspace*{2mm} & (d) & Female friendly spaces:  \\
&& $Q=1.8\times10^{-3}$ &&& $Q=1.2\times10^{-3}$ &&& $Q=1.6\times10^{-3}$ &&& $Q=8.2\times10^{-4}$\\
&& $I_{s}^{2}=0.47$ &&& $I_{s}^{2}=0.46$ &&& $I_{s}^{2}=0.28$ &&& $I_{s}^{2}=0.47$\\[5mm]
\multicolumn{3}{l}{\includegraphics[width=0.225\linewidth]{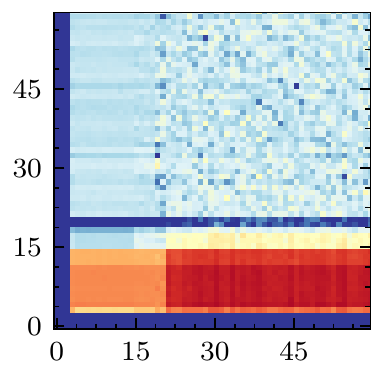}} &
\multicolumn{3}{l}{\includegraphics[width=0.225\linewidth]{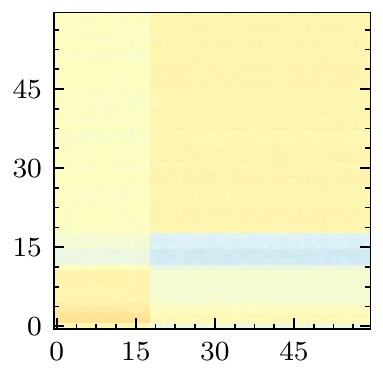}} &
\multicolumn{3}{l}{\includegraphics[width=0.225\linewidth]{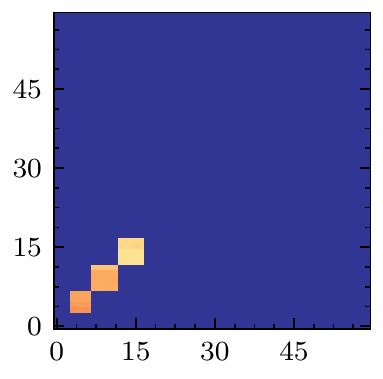}} &
\multicolumn{3}{l}{\includegraphics[width=0.225\linewidth]{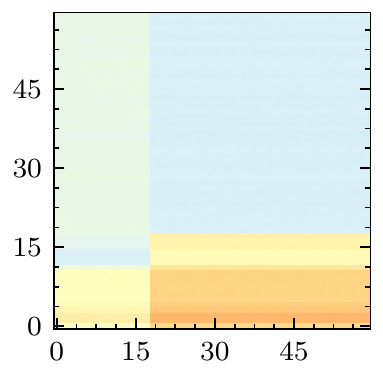}}\\[-1mm]
\hspace*{2mm} & (e) & Learning centres: &
\hspace*{2mm} & (f) & Non-food distribution & 
\hspace*{2mm} & (g) & Play groups:  &
\hspace*{2mm} & (h) & Pumps and latrines:  \\
&& $Q=-1.1\times10^{-3}$ &&& $Q=1.4\times10^{-3}$ &&& $Q=0.020$ &&& $Q=-2.4\times10^{-3}$\\
&& $I_{s}^{2}=0.71$ &&& $I_{s}^{2}=0.46$ &&& $I_{s}^{2}=0.056$ &&& $I_{s}^{2}=0.62$\\[5mm]
&&&
\multicolumn{3}{l}{\includegraphics[width=0.225\linewidth]{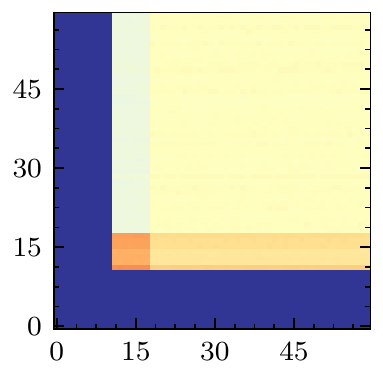}} &
\multicolumn{3}{l}{\includegraphics[width=0.225\linewidth]{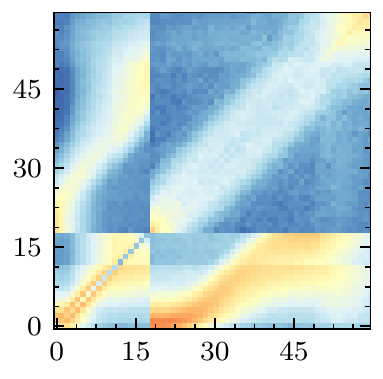}}\\[-1mm]
&&&
\hspace*{2mm} & (i) & Religious centres: &
\hspace*{2mm} & (j) & Shelters: & \\
&&&&& $Q=2.3\times10^{-3}$ &&& $Q=7.5\times10^{-3}$ \\
&&&&& $I_{s}^{2}=0.44$ &&& $I_{s}^{2}=0.38$ \\[5mm]
\end{tabular}
\parbox{0.8\linewidth}{\caption{\label{fig:UNCM_R_syoa}The reciprocal normalised contact matrices (UNCM$_\textbf{R}$)  by age as simulated in \JUNECOX. Note that the data inputs in (i) and (j) stem from a previous survey.}}
\end{figure}
Once we have determined the UNCM and confirmed that their stochastic uncertainties are within the uncertainties of the input interaction matrices, we can perform any custom binning for arbitrary group characteristics. 
Fig.~\ref{fig:UNCM_R_syoa} shows the final fully dis-aggregated (by age and venue) set of matrices for the Cox's Bazar refugee settlement based on the input contact matrices from the lightweight survey, combined with our highly-detailed agent-based model of the settlement. 
The combination of these two techniques leads to interesting consequences in the structure of the derived contact matrices. Contact rates from the light-weight survey provide the baseline coarse social interaction patterns between broad subgroups at a given venue. Whereas, the agent-based model embeds the dynamics from data on the social behavior of individuals, connecting many independent venues within the model.
In particular, we see bands due mainly to 11-18 year olds for two reasons.
Firstly, many behavioural patterns are defined differently for adults and children leading to attendance differences at 18.
Secondly, at 11 years of age men are permitted to attend the religious centres.
Due to the high rate of attendance observed at the religious centres, there is a drop in attendance at other non-religious centre venues of this age group relative to other age groups.
The corresponding UNCM and UNCM$_{\textbf{V}}$ can be found in Appendix~\ref{app:CM}. 

In Fig. \ref{fig:UNCM_R_syoa}, we can clearly see the effects of the different age groups and guiding contact rates. For example, we observe large differences in the number of contacts between all age groups with adults in the community centres relative to the distribution centres, with substructures based on the age profile of children attending these locations shown through the higher number of contacts in younger age brackets. In addition, the learning centre matrices show a clear mix of contacts between children in their mixed classes and their teachers - this matrix also encodes information on the enrollment rate of children in the education system, with lower enrollment rates as the age of children increase. Finally, the detailed information available on household and shelter composition appears in the shelter contact matrix which contains a number interesting features. 
We reconstruct a strong leading diagonal which represents persons of similar ages living together; siblings, parents and grandparents of similar ages the width of the band reflects spousal age gaps and minimal age gaps between consecutive siblings. 
Using more detailed information about the average age of parents at the birth of their first child we also develop off-diagonal structure in the upper left and lower right quadrants. There exists an almost linear structure corresponding to children and parents interacting and aging together. This structure then tapers off indicating interactions in multi-generational households before many children would leave home at around 18. The details of the household construction and the statistics that define it can be found in Appendix~\ref{app:DemographicProperties}. 

A simpler approach is to just assume that everyone contacts everyone else in these dense settings in the absence of other information - we also present the results for the corresponding UNCM$_V$ in Appendix \ref{app:CM}, Figure~\ref{fig:UNCM_V_syoa}. However, clearly there is a significant loss of information in doing this, in comparison to the mixed-method approach, as can be seen in the absence of structural detail in many of the UNCM$_V$ matrices.

\begin{figure}[h!]
\centering
\includegraphics[width=.9\linewidth]{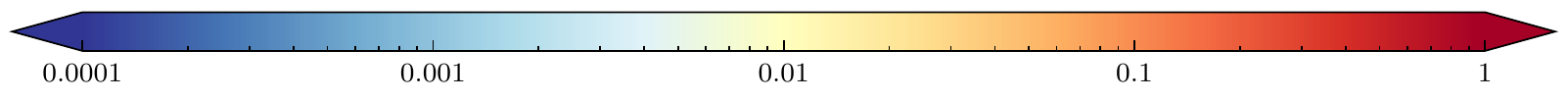}
\begin{tabular}{llllllllllll}
\multicolumn{3}{l}{\includegraphics[width=0.225\linewidth]{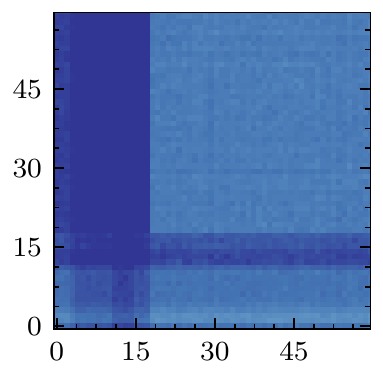}} &
\multicolumn{3}{l}{\includegraphics[width=0.225\linewidth]{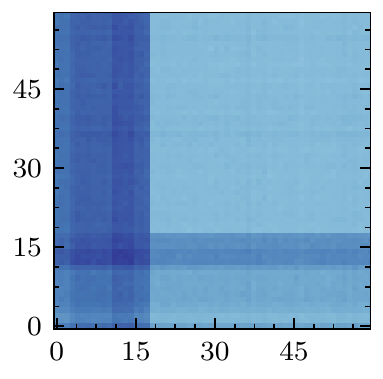}} &
\multicolumn{3}{l}{\includegraphics[width=0.225\linewidth]{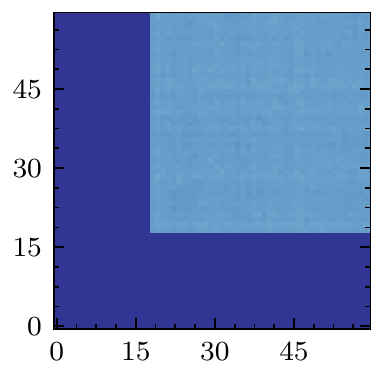}} &
\multicolumn{3}{l}{\includegraphics[width=0.225\linewidth]{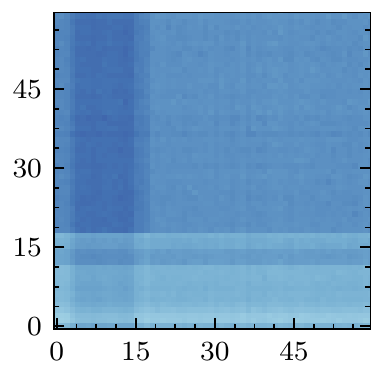}}\\[-1mm]
\hspace*{0mm} & (a) & Community centres: &
\hspace*{0mm} & (b) & Distribution centres:  &
\hspace*{0mm} & (c) & e-voucher outlets:  &
\hspace*{0mm} & (d) & Female friendly spaces:  \\
&& $Q=1.7\times10^{-3}$ &&& $Q=1.5\times10^{-3}$ &&& $Q=1.1\times10^{-3}$ &&& $Q=1.9\times10^{-3}$\\
&& $I_{s}^{2}=0.42$ &&& $I_{s}^{2}=0.42$ &&& $I_{s}^{2}=0.26$ &&& $I_{s}^{2}=0.42$\\[5mm]
\multicolumn{3}{l}{\includegraphics[width=0.225\linewidth]{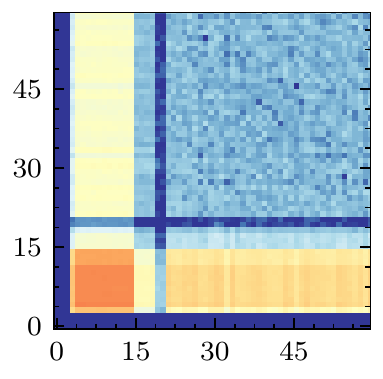}} &
\multicolumn{3}{l}{\includegraphics[width=0.225\linewidth]{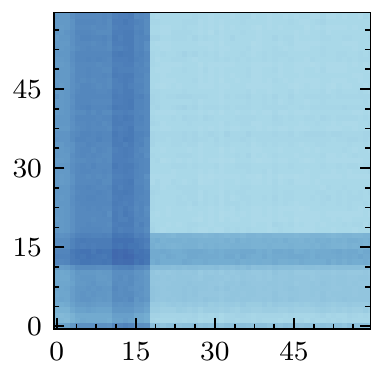}} &
\multicolumn{3}{l}{\includegraphics[width=0.225\linewidth]{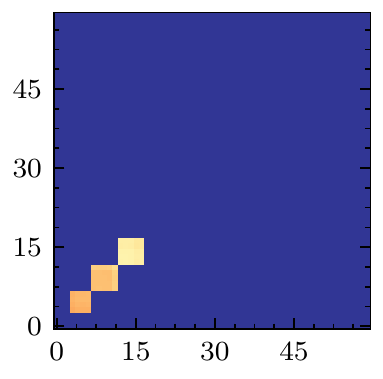}} &
\multicolumn{3}{l}{\includegraphics[width=0.225\linewidth]{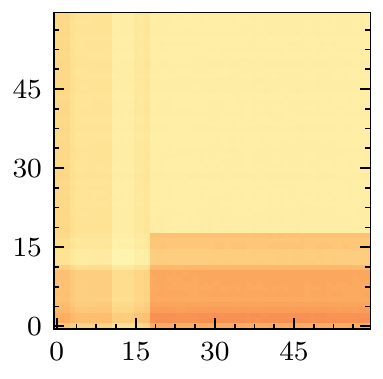}}\\[-1mm]
\hspace*{0mm} & (e) & Learning centres: &
\hspace*{0mm} & (f) & Non-food distribution & 
\hspace*{0mm} & (g) & Play groups:  &
\hspace*{0mm} & (h) & Pumps and latrines:  \\
&& $Q=9.5\times10^{-4}$ &&& $Q=1.7\times10^{-3}$ &&& $Q=2.0\times10^{-2}$ &&& $Q=-1.9\times10^{-4}$\\
&& $I_{s}^{2}=0.15$ &&& $I_{s}^{2}=0.37$ &&& $I_{s}^{2}=0.056$ &&& $I_{s}^{2}=5.3\times10^{-2}$\\[5mm]
&&&
\multicolumn{3}{l}{\includegraphics[width=0.225\linewidth]{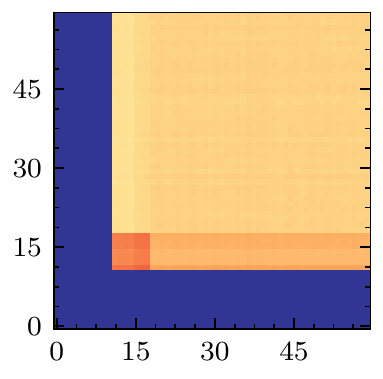}} &
\multicolumn{3}{l}{\includegraphics[width=0.225\linewidth]{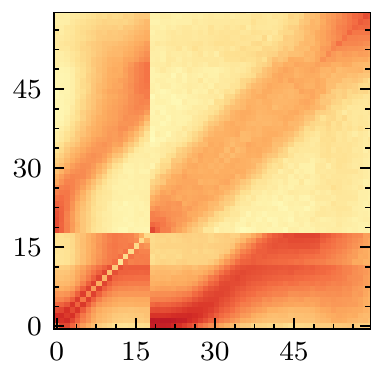}}\\[-1mm]
&&&
\hspace*{0mm} & (i) & Religious centres: &
\hspace*{0mm} & (j) & Shelters: & \\
&&&&& $Q=2.3\times10^{-3}$ &&& $Q=7.8\times10^{-3}$ \\
&&&&& $I_{s}^{2}=0.40$ &&& $I_{s}^{2}=0.37$ \\[5mm]
\end{tabular}
\parbox{0.8\linewidth}{\caption{\label{fig:PNCM_R_syoa}The population normalised contact matrices (PNCM$_\textbf{R}$) by age as simulated in \JUNECOX.}}
\end{figure}

Population normalised matrices can be calculated from the user normalised matrices with a simple re-scaling as described above. We present these for completeness and the varied utility of each normalisation in different model types in Appendix \ref{app:CM}, Figures~\ref{fig:PNCM_syoa},~\ref{fig:PNCM_R_syoa},~\ref{fig:PNCM_V_syoa}.

\section*{Discussion}
To the best of our knowledge, the matrices presented in this paper are the first contact matrices derived for a refugee settlement. While not collected using traditional survey methods, we use a mixed-method approach for their calculation, which presents a new way to collect contact data. This is particularly useful in settings, such as in refugee settlements, in which data collection can present many challenges, and therefore needs to be lightweight and integrated in to existing data collection regimes and programming.

We are able to perform closure tests on the contact matrices we derive and show that they clearly demonstrate great potential for a lightweight survey and an agent-based model to provide deeper insights into social environments when combined together. The survey and \JuneCox derived contact matrices are initially validated by a comparison of their Canberra distances over the survey subgroups $ij$. These Canberra distances are found to be very close to zero with the exception of the e-voucher outlets in which child - child contacts are higher in \JuneCox than reality. This discrepancy can be explained by considering that the survey has a high uncertainty in the expected child - child contacts, an error in which \JuneCox incorporates into the contact tracking algorithm.  

Further validation is performed with \JUNEUK derived matrices on age-disaggreagated contact matrices in which we are able to use other statistics such as $I^2_S$ and $Q$. These matrices were found to be in good agreement with other more intensive contact surveys. This validation ensures that the combination of coarse input contact matrices and the attendance rates responsible for agent dynamics yield representative contact patterns over all ages. 

In the case of refugee settlements, the derived contact matrices can be used to understand the social contact patterns using data already collected regularly by international organisations such as UNHCR, while being supplemented by data which can easily be collected by enumerators in a resource-efficient way. The highly-detailed matrices derived for the Cox's Bazar settlement demonstrate clear inter-age mixing patters which are crucial inputs to other epidemic models to represent realistic social mixing patterns. In particular, clear features are present in the matrices due to differing attendance rates and household compositions.

From the technical perspective, there are several further considerations and limitations to this methodology that become apparent when analysing the full age-disaggregated contact matrices (Fig. \ref{fig:UNCM_R_syoa} and Appendix \ref{app:CM}). These pertain to the way in which the data is collected and the model is constructed, and can be used as ways to diagnose the performance of the method:
\begin{enumerate}
    \item \textbf{Subgroup classification:}\\
    Subgroup classification refers to the broad definition of subgroups defined in the model. Throughout \JuneCox and \JUNEUK we define "Adults", "Children", "Teachers", "Workers" etc. which all have unique parameters and rules governing their behaviour.
    Subgroups defined by age can lead to strong banding artifacts in the contact matrices.
    These effects can be mitigated by blurring the age cut-off with some finite probability - \textit{e.g.} that a child of 17 may behave like an adult. 
    This mitigation should only be implemented in situations in which we are certain that there should not be a discontinuity in behaviours in the real world. 
    For example, only over 11 year old men are permitted to attend the religious centres and hence we expect a cut off in the contact matrices whereas in many other venues we expect a gradual shift in behaviour as children move into adolescence and then adulthood. 
    This can be a positive feature of the model - i.e. that the model represents the behavioural and movements patterns correctly and forces agents to make a choice between activities they perform as they would in real life - however, this relies on reasonable behavioural data, insights and assumptions.
    This is demonstrated most clearly in the \textit{shelters} contact matrices in which the household clustering places adults and children differently based on fixed rules derived from survey and census data (see Appendix \ref{app:DemographicProperties}).
    \item \textbf{Virtual venue demography:}\\
    The dynamics of virtual spaces in the simulation are dictated by the probabilistic attendance rates (see Figure.~\ref{fig:AgeProbabilities}) and age cut offs. The attendance rates are a function of age, sex, time and venue which leads to different demographies across the virtual spaces and therefore different social mixing behaviours.
    Again, due to the nature of the simulation in which we have strict probabilistic rules which determine the attendance of different subgroups (children, adults, age or sex etc.), we can get strong divisions between groupings. This is shown by the discontinuities in the heat-map representation of the contact matrices.
    In particular, only men over the age of 11 are permitted to attend the religious centres leading to discontinuities in the religious center contact matrices.
    In \JUNEUK, there is no simulation of parent-teacher interactions at school that might occur during pick up or drop off times, and the virtual school setting is strictly modelling student-teacher interactions where any teacher-teacher interactions would be restricted to the classroom setting. Further, no children attend any work place settings, and agents can only be employed or attend a work place venue between the ages of 18-65. The contact matrices produced from \JUNEUK therefore lack certain features shown by the BBC Pandemic project. However, this is a problem all such approaches that rely on an imperfect virtual representation of reality can experience.
    \item \textbf{Virtual world rules and behaviour patterns:}\\
    The combination of the above points leads to complex inter-connected behaviours across the simulation. 
    Considering the behaviour of coarser subgroups across all venues we see more general behaviours emerge; for instance, children are less likely to attend any virtual venue than adults, and men are more likely to attend any venue than women due to the attendance at religious centres which increases the overall rate of men not staying in the shelters compared to women, leading to an asymmetry in the shelter contact matrix.
    An 11-18 year old is more likely to see a 6-11 year old than the converse. 
    A 6-11 year old is more likely to be home than a 11-18 year old therefore on average in any timestep a 6-11 year old will not contact an 11-18 year old in shelters, but when the 11-18 year old is home they will likely contact the 6-11 year old.
    The normalisation of contacts by users (or population) and contact duration (as done throughout) makes this effect visible.
    There are other instances, such as the community centres, in which we see a banding effect which is an induced artifact from the movement criterion of the agents in the model (see Figure \ref{fig:AgeProbabilities}). The high attendance rate expected of 11+ men leads to a reduction in attendance of this group across all other venues, and many of the contact matrices show a banding effect between 11 - 18 due to this behaviour. 
\end{enumerate}

Given the level of detail contained within the model-derived contact matrices, they have the ability to reveal potential short-comings in both the survey setup as well as the modeling of the virtual world, as they reflect how sophisticated and well understood each venue type is. This means that the amount of resources needed to be expended on collecting more data on certain locations can be estimated in order to improve certain matrices. These can be traded-off against the resources available and the relative expected gain from their expenditure.
In this work, we validated our contact tracker in two very different models, \JUNEUK and \JuneCox. 
In the former, we demonstrated that the NCM$_\textbf{V}$ and NCM$_\textbf{R}$ agree well with data collected using traditional methods, {\it cf.} Tables~\ref{tab:CMStatistics} and ~\ref{tab:CompanySchoolUK}.
In the latter, NCM$_\textbf{V}$ type contact patterns are not available as our extracted contact matrices used coarse survey information on venue attendance to inform the simulation of contact patterns there, with the notable exception of the shelters, which are relatively precisely captured by the census data. 
Our mixed-method approach allows us to partially compensate for the gaps in detailed understanding of demographic structures at the lesser-known venues.
\section*{Conclusion}
In this work we demonstrate the complementary power of a lightweight contact survey, approximate details about venues and their attendance rates by different demographic groups, and an agent-based model to generate detailed social contact matrices. 
In the case of the Cox's Bazar refugee settlement, we use an existing model of the settlement developed using the \JUNE framework to perform a virtual contact survey, which is informed by the highly aggregate real world survey, to produce more granular contact matrices which can be further interrogated. 
Our constructed contact matrices will provide an important input to future disease spread modelling or social dynamic studies in the settlement, and provide a baseline which can be translated to other settlements as well.
Further, our method can easily be adapted to other settings for which detailed contact matrices are not available, thereby enabling the use of disease models in contexts where previously large assumptions would have had to have been made about contact patterns. Contact matrices form the backbone of many disease models, and so calculating them at a global scale, with the specific inclusion of those groups who are often most vulnerable to disease spread, is essential~\cite{aylett-bullock_epidemiological_2022}.
\section*{Code and Availability}
\begin{itemize}
    \item \JUNE and \textbf{\JUNEUK}: The current public release of the \JUNE simulation framework, and by extension the latest version of the \JUNEUK model, can be found at \url{https://github.com/IDAS-Durham/JUNE}
    \item \textbf{\JuneCox}: The current public release of \JuneCox epidemic model can be found at \url{https://github.com/UNGlobalPulse/UNGP-settlement-modelling}
    \item Data: The data from the survey is available by application at \url{https://microdata.unhcr.org/index.php/catalog/587}
    \item Contact Survey: Details and calculation at
    \url{https://github.com/UNGlobalPulse/UNGP-contact-survey}
    \item \JUNEUK Household contact matrix: Details and calculation at \url{https://github.com/IDAS-Durham/june_household_matrix_calculation}
    \item Contact Matrix Results: Our contact matrices are reported here \url{https://github.com/IDAS-Durham/june_mixed_method_CM_results} formatted in excel documents for convenience.
\end{itemize}

\bibliography{contact_matrices, references}

%

\section*{Ethics}

The survey run as part of this study was approved by the ethics committee of Durham University, reference: PHYS-2020-09-04T10:24:22-gnvq71.

\section*{Acknowledgements}
We would like to thank the UNHCR Cox's Bazar teams for their helpful comments on this work and for the support in setting up and running the survey in the settlement. In particular we would like to thank Hussien Ahmad, Hosna Ara Begum, Mahfuzur Rahman, and all the all the members of the Information Management, Community Based Protection and Public Health teams. The authors would also like to thank Giulia Zarpellon and Miguel Luengo-Oroz for their helpful comments and suggestions throughout this work. United Nations Global Pulse work is supported by the Governments of Sweden and Canada, and the William and Flora Hewlett Foundation. JA-B, DS and JW were supported by the Centre for Doctoral Training in Data Intensive Science under UKRI-STFC grant number ST/P006744/1 for parts of this work. DS is also funded by STFC through a Data Innovation Fellowship (ST/R005516/1). 
JW and DS received financial support through the EPSRC IAA project "{\it Creating humanitarian impact through data modelling: Collaborating with WHO and UN Global Pulse}".
FK gratefully acknowledges funding as Royal Society Wolfson Research fellow. 
This work used the DiRAC@Durham facility managed by the Institute for Computational Cosmology on behalf of the STFC DiRAC HPC Facility (www.dirac.ac.uk). The equipment was funded by BEIS capital funding via STFC capital grants ST/K00042X/1, ST/P002293/1, ST/R002371/1 and ST/S002502/1, Durham University and STFC operations grant ST/R000832/1. DiRAC is part of the National e-Infrastructure.

\section*{Author contributions statement}

Study conception: JA-B; data collection: JA-B, JW; running simulations: JW; analysis of results: JW, DS, JA-B, FK; interpretation of results: JW, JA-B, DS, FK, AGKM, ESE, SH; draft manuscript preparation: JW, JA-B, FK. All authors reviewed the results and approved the final version of the manuscript.

\section*{Disclaimer}

The authors alone are responsible for the views expressed in this article and they do not necessarily represent the views, decisions or policies of the institutions with which they are affiliated including the United Nations.

\newpage
\appendix
\renewcommand{\tabcolsep}{3pt}
\clearpage
\section{Contact Matrices}\label{app:CM}
Here we present the remaining contact matrices derived from \JUNECOX\footnote{We interpret contact matrix $\Delta_{ij}$ such that person $i$ contacts person $j$ and graphically as subgroup on $x$-axis contacts subgroup on $y$-axis.}.
\subsection{UNCM$_\textbf{V}$ Interaction}
\begin{figure}[h!]
\centering
\includegraphics[width=.9\linewidth]{figures/colourbar_UNCM_Interaction.pdf}
\begin{tabular}{llllllllllll}
\multicolumn{3}{l}{\includegraphics[width=0.225\linewidth]{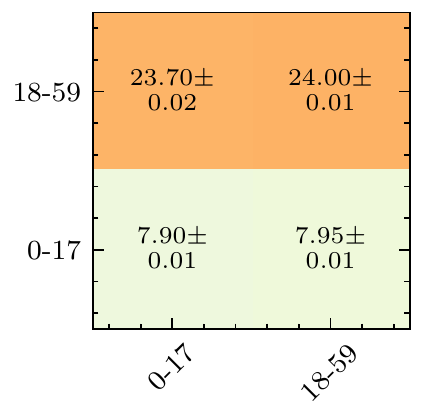}} &
\multicolumn{3}{l}{\includegraphics[width=0.225\linewidth]{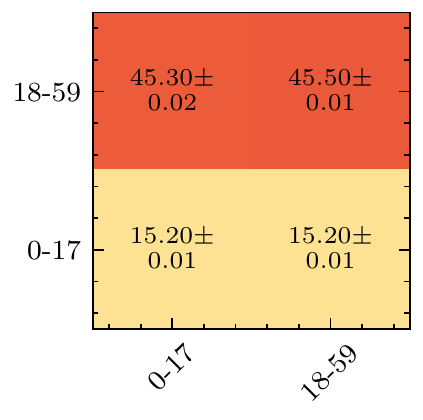}} &
\multicolumn{3}{l}{\includegraphics[width=0.225\linewidth]{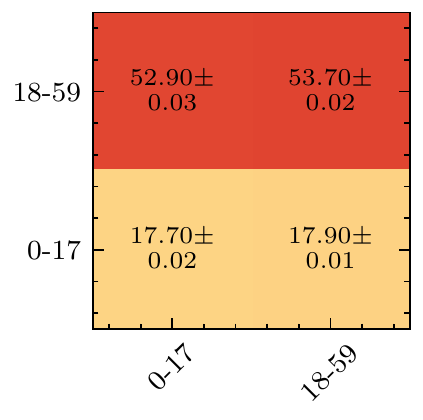}} &
\multicolumn{3}{l}{\includegraphics[width=0.225\linewidth]{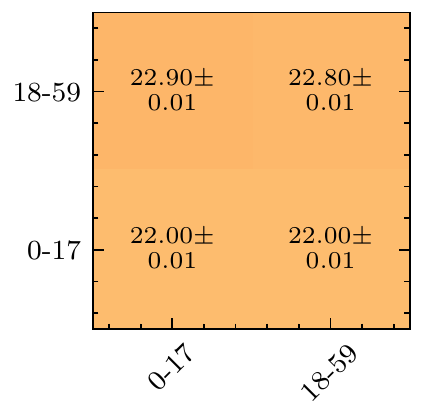}}\\[-1mm]
\hspace*{4mm} & (a) & Community centres: &
\hspace*{4mm} & (b) & Distribution centres:  &
\hspace*{4mm} & (c) & e-voucher outlets:  &
\hspace*{4mm} & (d) & Female friendly spaces:  \\
&& $D_C = 0.49$ &&& $D_C =  0.66$ &&& $D_C =  0.88$ &&& $D_C =  0.64$\\[5mm]
\multicolumn{3}{l}{\includegraphics[width=0.225\linewidth]{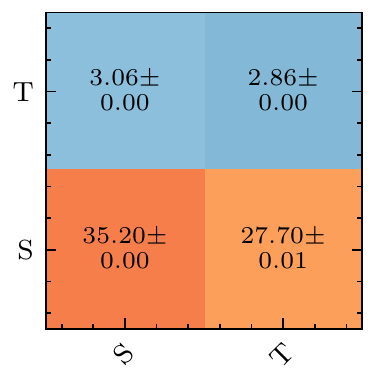}} &
\multicolumn{3}{l}{\includegraphics[width=0.225\linewidth]{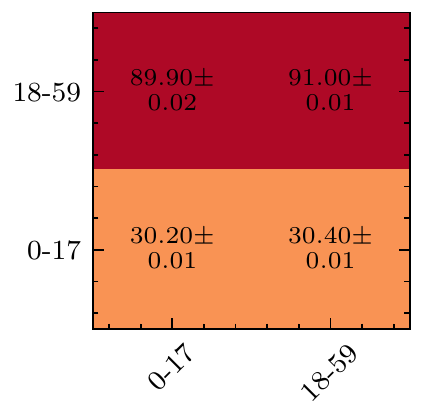}} &
\multicolumn{3}{l}{\includegraphics[width=0.225\linewidth]{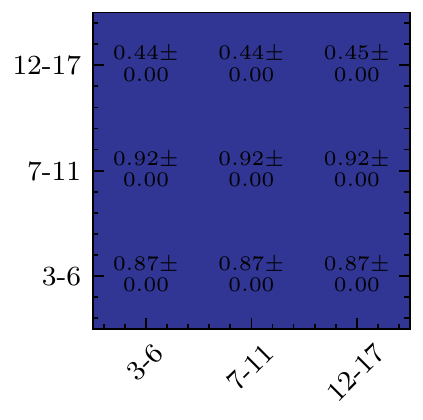}} &
\multicolumn{3}{l}{\includegraphics[width=0.225\linewidth]{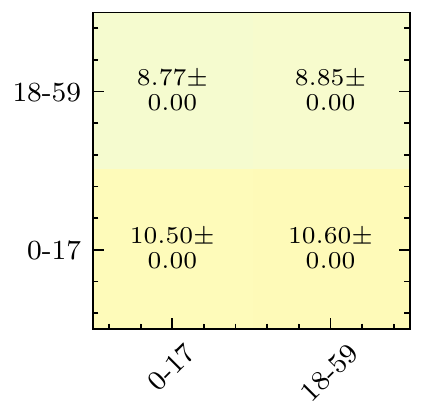}}\\[-1mm]
\hspace*{4mm} & (e) & Learning centres: &
\hspace*{4mm} & (f) & Non-food distribution & 
\hspace*{4mm} & (g) & Play groups:  &
\hspace*{4mm} & (h) & Pumps and latrines:  \\
&& $D_C =  0.45$ &&& centres: $D_C =  0.81$ &&& $D_C =  0.73$ &&& $D_C =  0.56$\\[5mm]
&&&
\multicolumn{3}{l}{\includegraphics[width=0.225\linewidth]{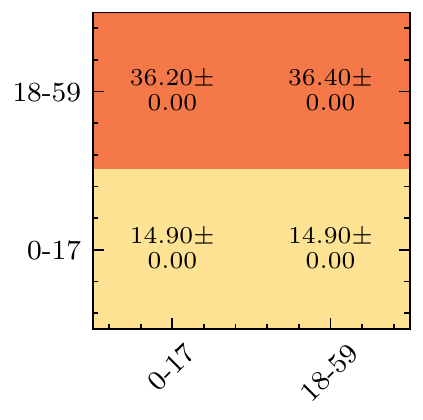}} &
\multicolumn{3}{l}{\includegraphics[width=0.225\linewidth]{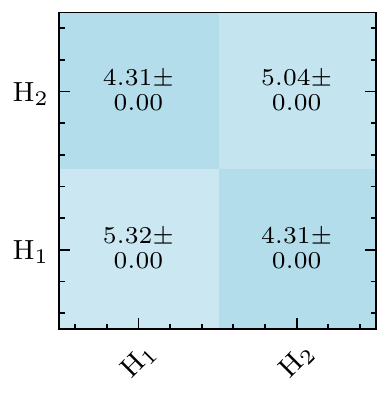}} &&&\\[-1mm]
&&&
\hspace*{4mm} & (i) & Religious centres: &
\hspace*{4mm} & (j) & Shelters: & \\
&&&&& $D_C =  0.64$ &&& $D_C =  0.19$ \\[5mm]
\end{tabular}
\parbox{0.8\linewidth}{\caption{\label{fig:UNCM_V_Interaction}The normalised venue contact matrices (UNCM$_\textbf{V}$) by input survey subgroups as simulated in \JUNECOX.}}
\end{figure}
\clearpage
\subsection{i) UNCM}
\begin{figure}[h!]
\centering
\includegraphics[width=.9\linewidth]{figures/colourbar_UNCM_R_syoa.pdf}
\begin{tabular}{llllllllllll}
\multicolumn{3}{l}{\includegraphics[width=0.225\linewidth]{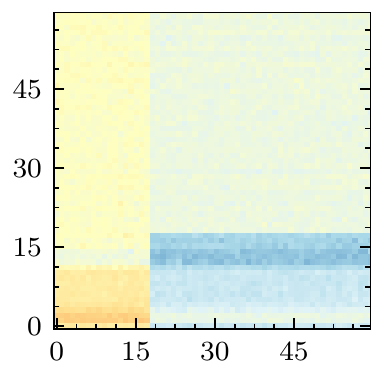}} &
\multicolumn{3}{l}{\includegraphics[width=0.225\linewidth]{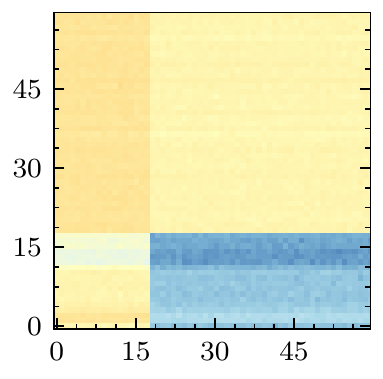}} &
\multicolumn{3}{l}{\includegraphics[width=0.225\linewidth]{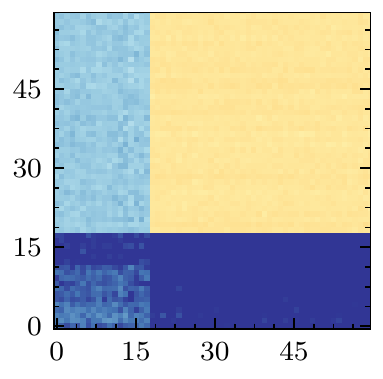}} &
\multicolumn{3}{l}{\includegraphics[width=0.225\linewidth]{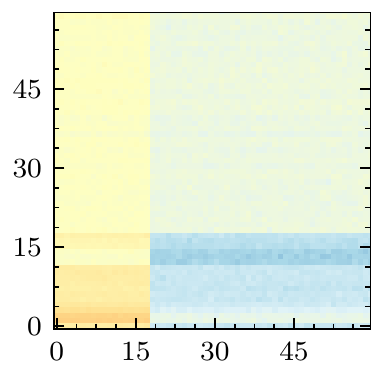}}\\[-1mm]
\hspace*{0mm} & (a) & Community centres: &
\hspace*{0mm} & (b) & Distribution centres:  &
\hspace*{0mm} & (c) & e-voucher outlets:  &
\hspace*{0mm} & (d) & Female friendly spaces:  \\
&& $Q=1.7\times10^{-3}$ &&& $Q=1.5\times10^{-3}$ &&& $Q=1.1\times10^{-3}$ &&& $Q=1.9\times10^{-3}$\\
&& $I_{s}^{2}=0.47$ &&& $I_{s}^{2}=0.46$ &&& $I_{s}^{2}=0.29$ &&& $I_{s}^{2}=0.41$\\[5mm]
\multicolumn{3}{l}{\includegraphics[width=0.225\linewidth]{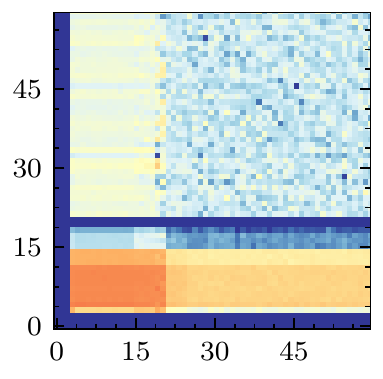}} &
\multicolumn{3}{l}{\includegraphics[width=0.225\linewidth]{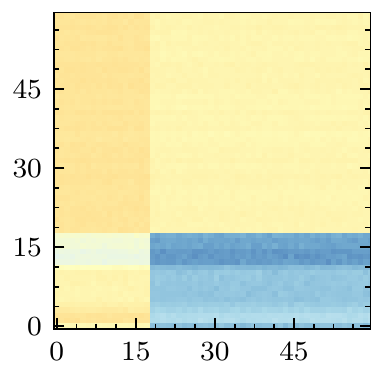}} &
\multicolumn{3}{l}{\includegraphics[width=0.225\linewidth]{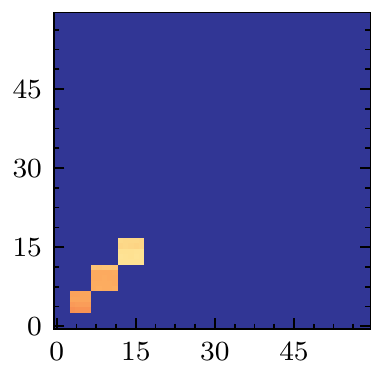}} &
\multicolumn{3}{l}{\includegraphics[width=0.225\linewidth]{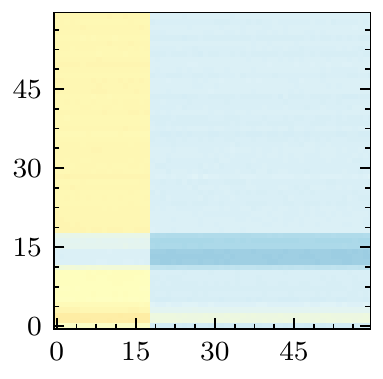}}\\[-1mm]
\hspace*{0mm} & (e) & Learning centres: &
\hspace*{0mm} & (f) & Non-food distribution & 
\hspace*{0mm} & (g) & Play groups:  &
\hspace*{0mm} & (h) & Pumps and latrines:  \\
&& $Q=9.4\times10^{-4}$ &&& $Q=1.7\times10^{-3}$ &&& $Q=2.0\times10^{-2}$ &&& $Q=-1.9\times10^{-4}$\\
&& $I_{s}^{2}=0.24$ &&& $I_{s}^{2}=0.46$ &&& $I_{s}^{2}=0.056$ &&& $I_{s}^{2}=0.51$\\[5mm]
&&&
\multicolumn{3}{l}{\includegraphics[width=0.225\linewidth]{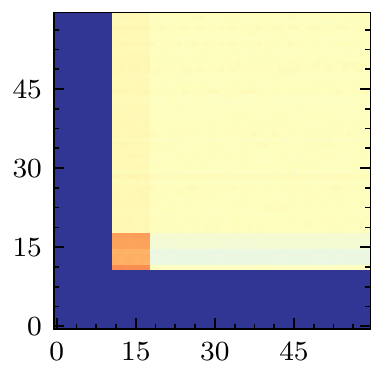}} &
\multicolumn{3}{l}{\includegraphics[width=0.225\linewidth]{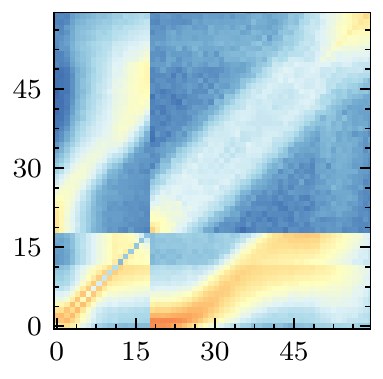}}\\[-1mm]
&&&
\hspace*{0mm} & (i) & Religious centres: &
\hspace*{0mm} & (j) & Shelters: & \\
&&&&& $Q=2.3\times10^{-3}$ &&& $Q=7.8\times10^{-3}$ \\
&&&&& $I_{s}^{2}=0.40$ &&& $I_{s}^{2}=0.37$ \\[5mm]
\end{tabular}
\parbox{0.8\linewidth}{\caption{\label{fig:UNCM_syoa}The normalised contact matrices (UNCM) by age as simulated in \JUNECOX. Note that the data inputs in (i) and (j) stem from a previous survey.}}
\end{figure}
\clearpage
\subsection{ii) UNCM$_\textbf{V}$}
\begin{figure}[h!]
\centering
\includegraphics[width=.9\linewidth]{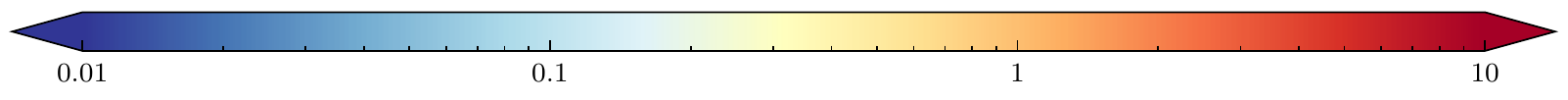}
\begin{tabular}{llllllllllll}
\multicolumn{3}{l}{\includegraphics[width=0.225\linewidth]{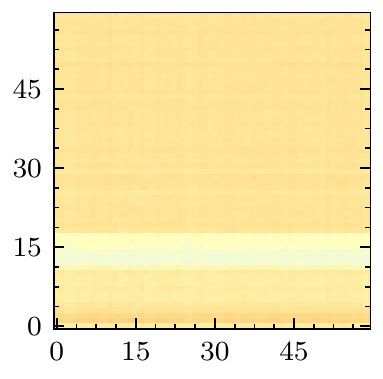}} &
\multicolumn{3}{l}{\includegraphics[width=0.225\linewidth]{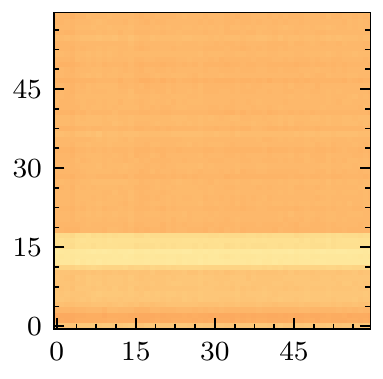}} &
\multicolumn{3}{l}{\includegraphics[width=0.225\linewidth]{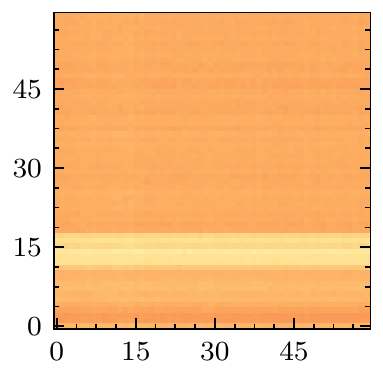}} &
\multicolumn{3}{l}{\includegraphics[width=0.225\linewidth]{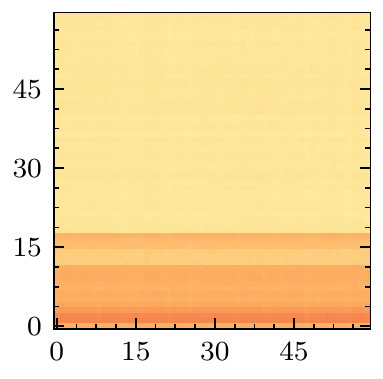}}\\[-1mm]
\hspace*{2mm} & (a) & Community centres: &
\hspace*{2mm} & (b) & Distribution centres:  &
\hspace*{2mm} & (c) & e-voucher outlets:  &
\hspace*{2mm} & (d) & Female friendly spaces:  \\
&& $Q=1.8\times10^{-6}$ &&& $Q=9.2\times10^{-6}$ &&& $Q=8.1\times10^{-5}$ &&& $Q=5.7\times10^{-5}$\\
&& $I_{s}^{2}=0.50$ &&& $I_{s}^{2}=0.50$ &&& $I_{s}^{2}=0.50$ &&& $I_{s}^{2}=0.50$\\[5mm]
\multicolumn{3}{l}{\includegraphics[width=0.225\linewidth]{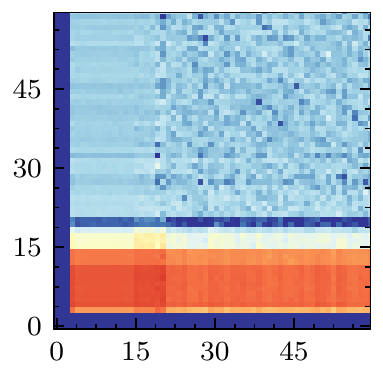}} &
\multicolumn{3}{l}{\includegraphics[width=0.225\linewidth]{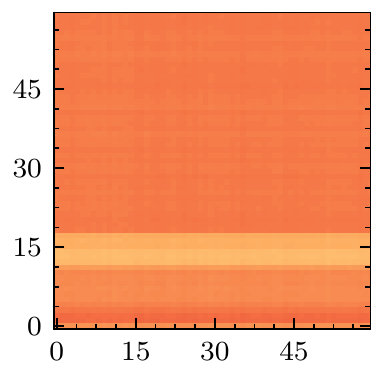}} &
\multicolumn{3}{l}{\includegraphics[width=0.225\linewidth]{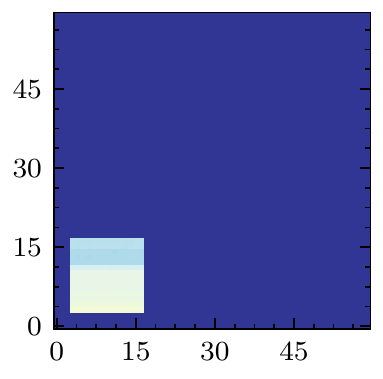}} &
\multicolumn{3}{l}{\includegraphics[width=0.225\linewidth]{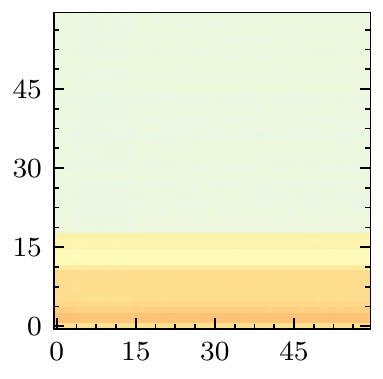}}\\[-1mm]
\hspace*{2mm} & (e) & Learning centres: &
\hspace*{2mm} & (f) & Non-food distribution & 
\hspace*{2mm} & (g) & Play groups:  &
\hspace*{2mm} & (h) & Pumps and latrines:  \\
&& $Q=4.1\times10^{-4}$ &&& $Q=3.2\times10^{-5}$ &&& $Q=3.6\times10^{-5}$ &&& $Q=5.6\times10^{-5}$\\
&& $I_{s}^{2}=0.28$ &&& $I_{s}^{2}=0.50$ &&& $I_{s}^{2}=0.48$ &&& $I_{s}^{2}=0.48$\\[5mm]
&&&
\multicolumn{3}{l}{\includegraphics[width=0.225\linewidth]{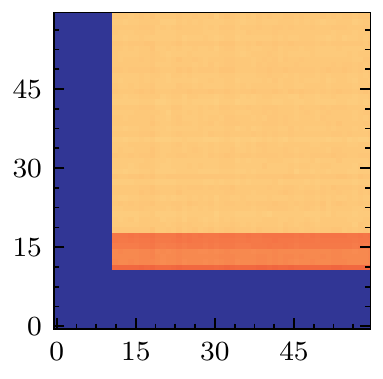}} &
\multicolumn{3}{l}{\includegraphics[width=0.225\linewidth]{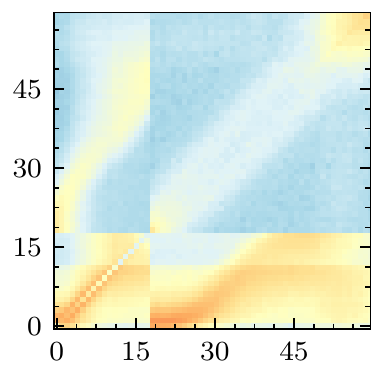}}\\[-1mm]
&&&
\hspace*{2mm} & (i) & Religious centres: &
\hspace*{2mm} & (j) & Shelters: & \\
&&&&& $Q=3.9\times10^{-6}$ &&& $Q=5.3\times10^{-3}$ \\
&&&&& $I_{s}^{2}=0.52$ &&& $I_{s}^{2}=0.41$ \\[5mm]
\end{tabular}
\parbox{0.8\linewidth}{\caption{\label{fig:UNCM_V_syoa}The normalised venue contact matrices (UNCM$_\textbf{V}$)  by age as simulated in \JUNECOX. Note that the data inputs in (i) and (j) stem from a previous survey.}}
\end{figure}
\clearpage
\subsection{iii) PNCM}
\begin{figure}[h!]
\centering
\includegraphics[width=.9\linewidth]{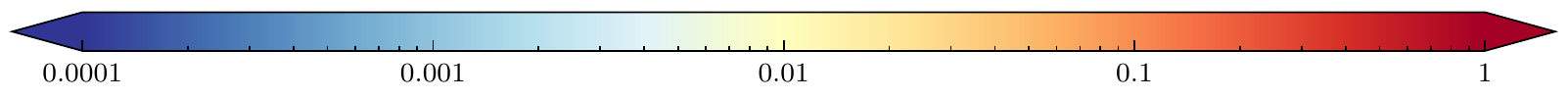}
\begin{tabular}{llllllllllll}
\multicolumn{3}{l}{\includegraphics[width=0.225\linewidth]{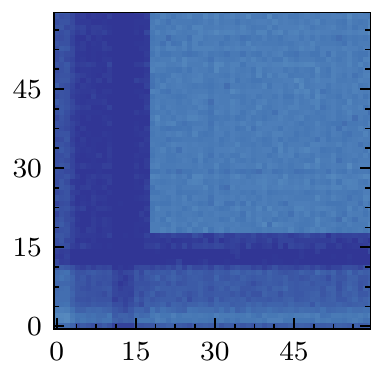}} &
\multicolumn{3}{l}{\includegraphics[width=0.225\linewidth]{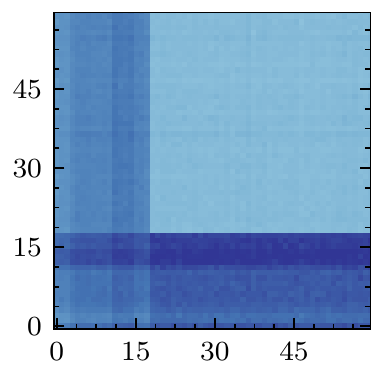}} &
\multicolumn{3}{l}{\includegraphics[width=0.225\linewidth]{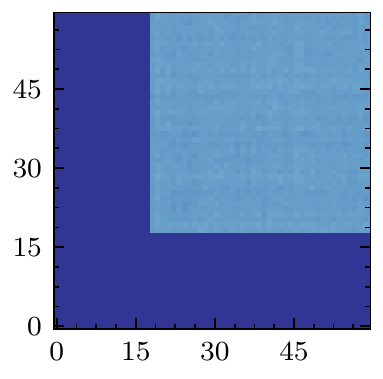}} &
\multicolumn{3}{l}{\includegraphics[width=0.225\linewidth]{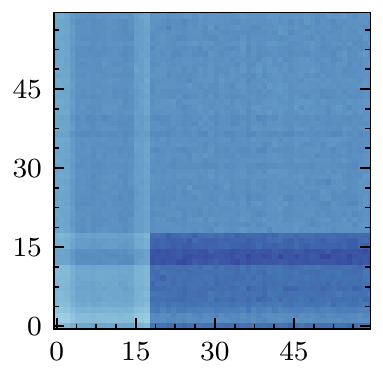}}\\[-1mm]
\hspace*{2mm} & (a) & Community centres: &
\hspace*{2mm} & (b) & Distribution centres:  &
\hspace*{2mm} & (c) & e-voucher outlets:  &
\hspace*{2mm} & (d) & Female friendly spaces:  \\
&& $Q=1.7\times10^{-3}$ &&& $Q=1.5\times 10^{-3}$ &&& $Q=1.1\times 10^{-3}$ &&& $Q=1.9\times 10^{-3}$\\
&& $I_{s}^{2}=0.42$ &&& $I_{s}^{2}=0.36$ &&& $I_{s}^{2}=0.26$ &&& $I_{s}^{2}=0.42$\\[5mm]
\multicolumn{3}{l}{\includegraphics[width=0.225\linewidth]{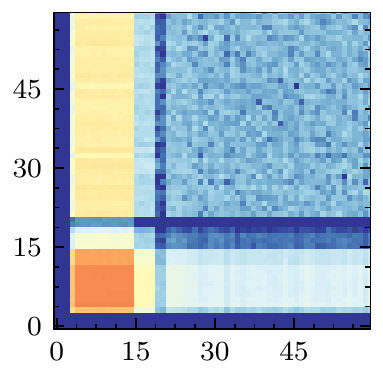}} &
\multicolumn{3}{l}{\includegraphics[width=0.225\linewidth]{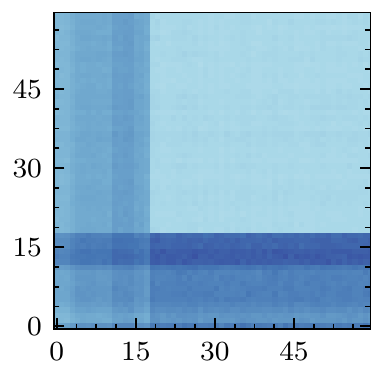}} &
\multicolumn{3}{l}{\includegraphics[width=0.225\linewidth]{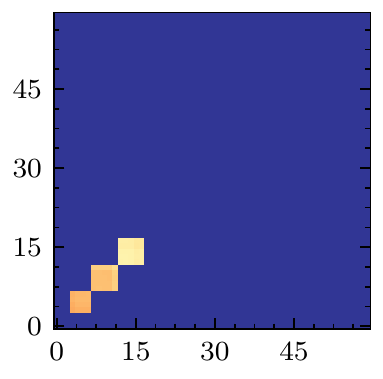}} &
\multicolumn{3}{l}{\includegraphics[width=0.225\linewidth]{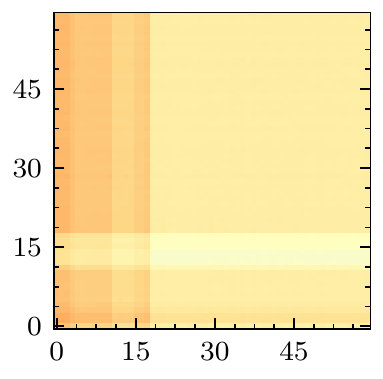}}\\[-1mm]
\hspace*{2mm} & (e) & Learning centres: &
\hspace*{2mm} & (f) & Non-food distribution & 
\hspace*{2mm} & (g) & Play groups:  &
\hspace*{2mm} & (h) & Pumps and latrines:  \\
&& $Q=9.4\times 10^{-4}$ &&& $Q=1.7\times 10^{-3}$ &&& $Q=2.0\times 10^{-2}$ &&& $Q=-1.9\times10^{-4}$\\
&& $I_{s}^{2}=0.15$ &&& $I_{s}^{2}=0.37$ &&& $I_{s}^{2}=5.3 \times 10^{-2}$ &&& $I_{s}^{2}=0.52$\\[5mm]
&&&
\multicolumn{3}{l}{\includegraphics[width=0.225\linewidth]{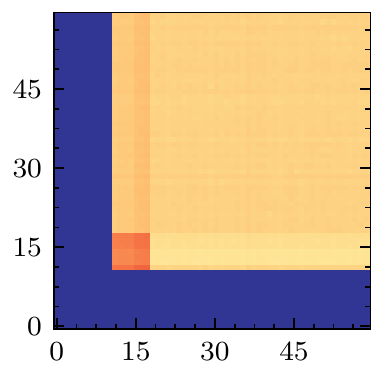}} &
\multicolumn{3}{l}{\includegraphics[width=0.225\linewidth]{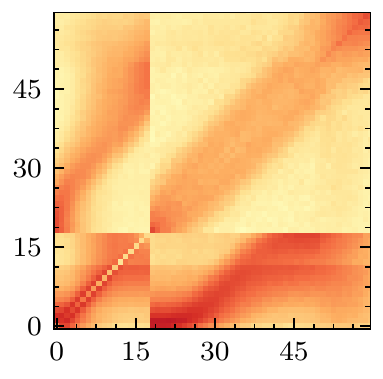}}\\[-1mm]
&&&
\hspace*{2mm} & (i) & Religious centres: &
\hspace*{2mm} & (j) & Shelters: & \\
&&&&& $Q=2.3\times 10^{-3}$ &&& $Q=7.8\times 10^{-3}$ \\
&&&&& $I_{s}^{2}=0.40$ &&& $I_{s}^{2}=0.38$ \\[5mm]
\end{tabular}
\parbox{0.8\linewidth}{\caption{\label{fig:PNCM_syoa}The normalised venue contact matrices (PNCM)  by age as simulated in \JUNECOX. Note that the data inputs in (i) and (j) stem from a previous survey.}}
\end{figure}
\clearpage
\subsection{iv) PNCM$_\textbf{R}$}
\begin{figure}[h!]
\centering
\includegraphics[width=.9\linewidth]{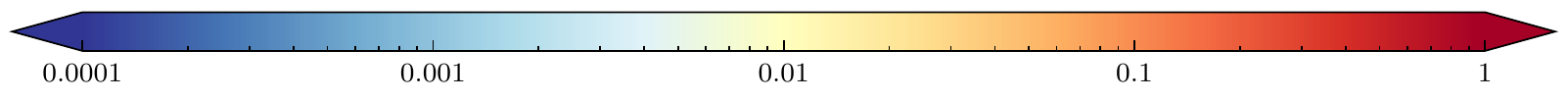}
\begin{tabular}{llllllllllll}
\multicolumn{3}{l}{\includegraphics[width=0.225\linewidth]{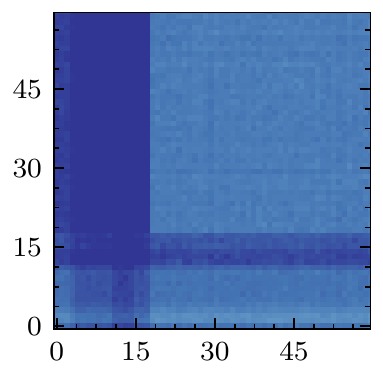}} &
\multicolumn{3}{l}{\includegraphics[width=0.225\linewidth]{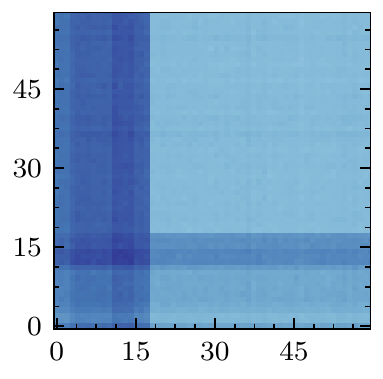}} &
\multicolumn{3}{l}{\includegraphics[width=0.225\linewidth]{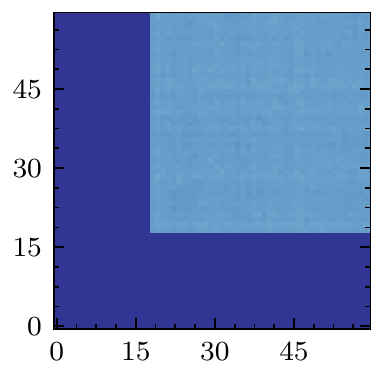}} &
\multicolumn{3}{l}{\includegraphics[width=0.225\linewidth]{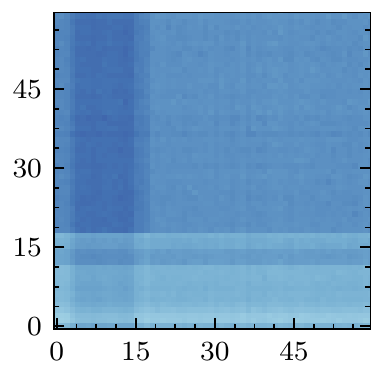}}\\[-1mm]
\hspace*{2mm} & (a) & Community centres: &
\hspace*{2mm} & (b) & Distribution centres:  &
\hspace*{2mm} & (c) & e-voucher outlets:  &
\hspace*{2mm} & (d) & Female friendly spaces:  \\
&& $Q=1.6\times10^{-3}$ &&& $Q=1.2\times10^{-3}$ &&& $Q=1.4\times10^{-3}$ &&& $Q=8.3\times10^{-4}$\\
&& $I_{s}^{2}=0.45$ &&& $I_{s}^{2}=0.42$ &&& $I_{s}^{2}=0.27$ &&& $I_{s}^{2}=0.50$\\[5mm]
\multicolumn{3}{l}{\includegraphics[width=0.225\linewidth]{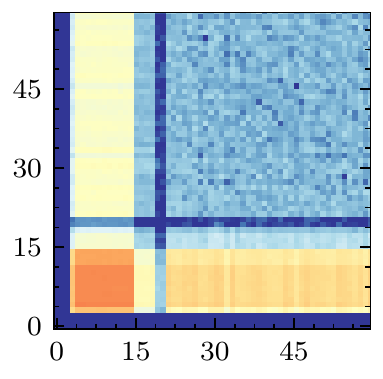}} &
\multicolumn{3}{l}{\includegraphics[width=0.225\linewidth]{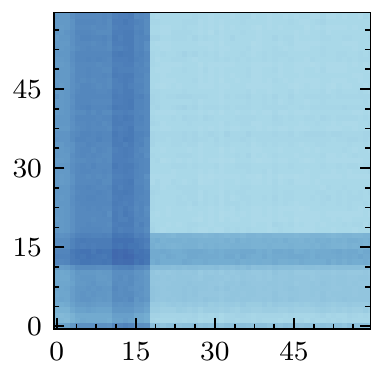}} &
\multicolumn{3}{l}{\includegraphics[width=0.225\linewidth]{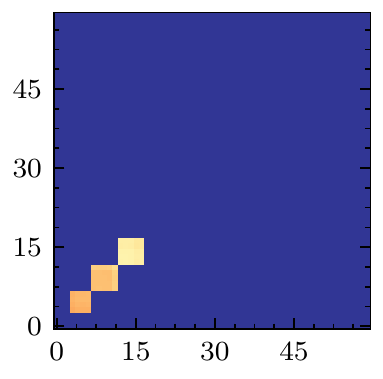}} &
\multicolumn{3}{l}{\includegraphics[width=0.225\linewidth]{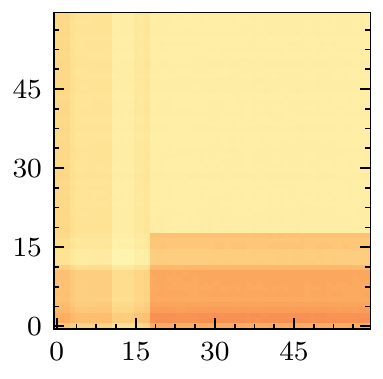}}\\[-1mm]
\hspace*{2mm} & (e) & Learning centres: &
\hspace*{2mm} & (f) & Non-food distribution & 
\hspace*{2mm} & (g) & Play groups:  &
\hspace*{2mm} & (h) & Pumps and latrines:  \\
&& $Q=1.1\times10^{-3}$ &&& $Q=1.3\times10^{-3}$ &&& $Q=2.0\times10^{-2}$ &&& $Q=-2.4\times10^{-3}$\\
&& $I_{s}^{2}=0.18$ &&& $I_{s}^{2}=0.43$ &&& $I_{s}^{2}=0.053$ &&& $I_{s}^{2}=0.62$\\[5mm]
&&&
\multicolumn{3}{l}{\includegraphics[width=0.225\linewidth]{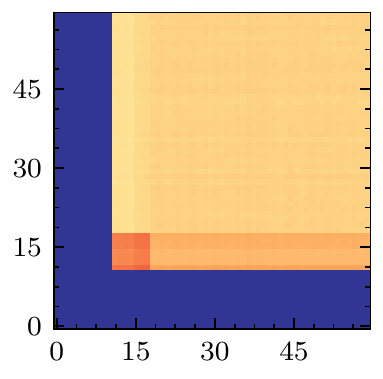}} &
\multicolumn{3}{l}{\includegraphics[width=0.225\linewidth]{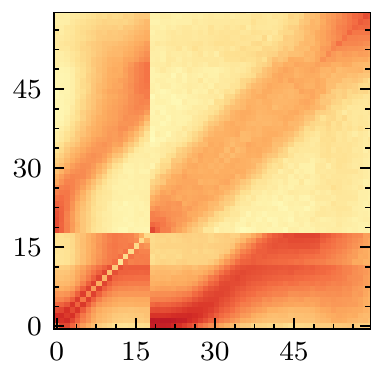}}\\[-1mm]
&&&
\hspace*{2mm} & (i) & Religious centres: &
\hspace*{2mm} & (j) & Shelters: & \\
&&&&& $Q=2.2\times10^{-3}$ &&& $Q=7.5\times10^{-3}$ \\
&&&&& $I_{s}^{2}=0.43$ &&& $I_{s}^{2}=0.39$ \\[5mm]
\end{tabular}
\parbox{0.8\linewidth}{\caption{\label{fig:PNCM_R_syoa}The normalised venue contact matrices (PNCM$_\textbf{R}$)  by age as simulated in \JUNECOX. Note that the data inputs in (i) and (j) stem from a previous survey.}}
\end{figure}
\clearpage
\subsection{v) PNCM$_\textbf{V}$}
\begin{figure}[h!]
\centering
\includegraphics[width=.9\linewidth]{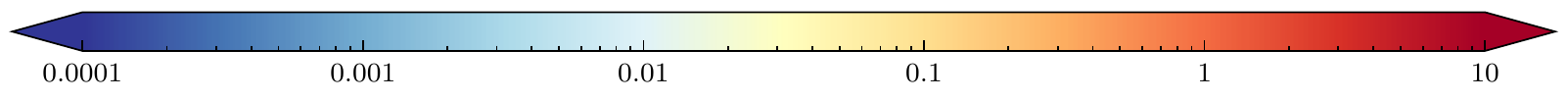}
\begin{tabular}{llllllllllll}
\multicolumn{3}{l}{\includegraphics[width=0.225\linewidth]{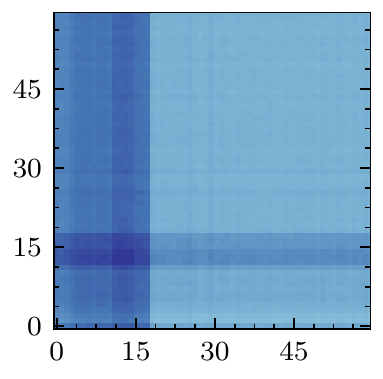}} &
\multicolumn{3}{l}{\includegraphics[width=0.225\linewidth]{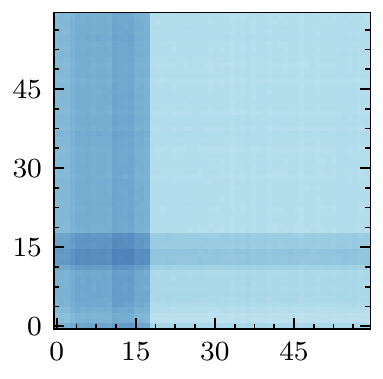}} &
\multicolumn{3}{l}{\includegraphics[width=0.225\linewidth]{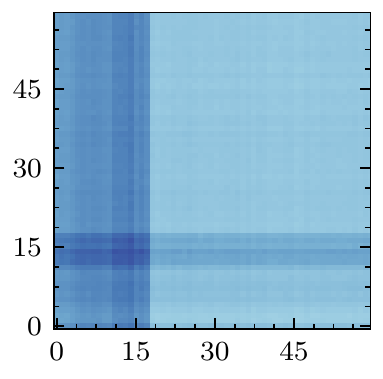}} &
\multicolumn{3}{l}{\includegraphics[width=0.225\linewidth]{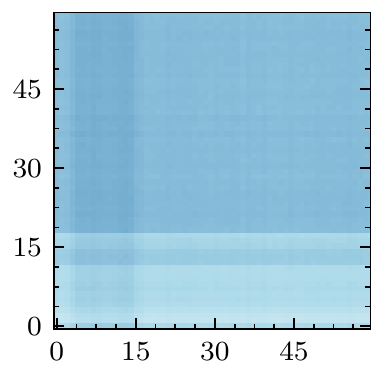}}\\[-1mm]
\hspace*{2mm} & (a) & Communal space: &
\hspace*{2mm} & (b) & Distribution centres:  &
\hspace*{2mm} & (c) & e-voucher outlets:  &
\hspace*{2mm} & (d) & Female friendly spaces:  \\
&& $Q=1.8\times10^{-6}$ &&& $Q=9.2\times10^{-6}$ &&& $Q=8.1\times10^{-5}$ &&& $Q=5.7\times10^{-5}$\\
&& $I_{s}^{2}=0.45$ &&& $I_{s}^{2}=0.45$ &&& $I_{s}^{2}=0.44$ &&& $I_{s}^{2}=0.52$\\[5mm]
\multicolumn{3}{l}{\includegraphics[width=0.225\linewidth]{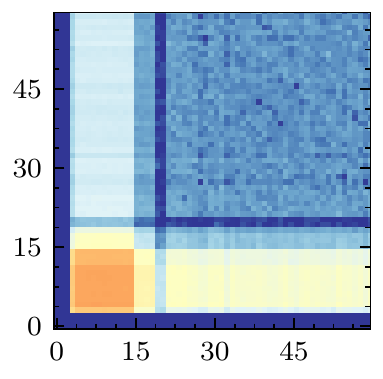}} &
\multicolumn{3}{l}{\includegraphics[width=0.225\linewidth]{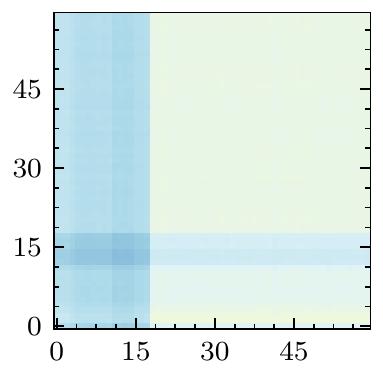}} &
\multicolumn{3}{l}{\includegraphics[width=0.225\linewidth]{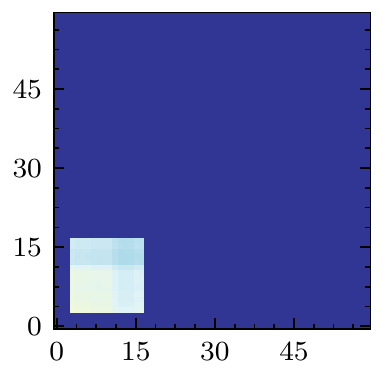}} &
\multicolumn{3}{l}{\includegraphics[width=0.225\linewidth]{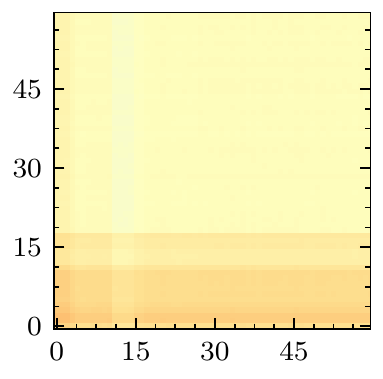}}\\[-1mm]
\hspace*{2mm} & (e) & Learning centres: &
\hspace*{2mm} & (f) & Non-food distribution & 
\hspace*{2mm} & (g) & Play groups:  &
\hspace*{2mm} & (h) & Pumps and latrines:  \\
&& $Q=4.1\times10^{-4}$ &&& $Q=3.1\times10^{-5}$ &&& $Q=3.6\times10^{-5}$ &&& $Q=5.7\times10^{-5}$\\
&& $I_{s}^{2}=0.13$ &&& $I_{s}^{2}=0.42$ &&& $I_{s}^{2}=0.44$ &&& $I_{s}^{2}=0.48$\\[5mm]
&&&
\multicolumn{3}{l}{\includegraphics[width=0.225\linewidth]{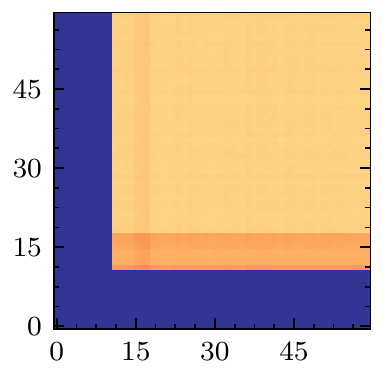}} &
\multicolumn{3}{l}{\includegraphics[width=0.225\linewidth]{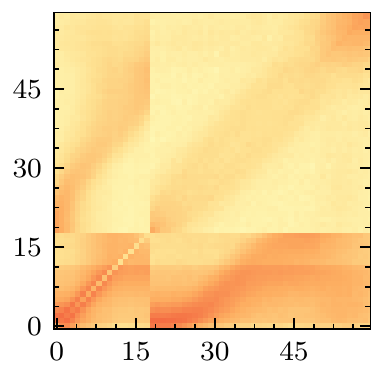}}\\[-1mm]
&&&
\hspace*{2mm} & (i) & Religious centres: &
\hspace*{2mm} & (j) & Shelters: & \\
&&&&& $Q=3.9\times10^{-6}$ &&& $Q=5.3\times10^{-3}$ \\
&&&&& $I_{s}^{2}=0.52$ &&& $I_{s}^{2}=0.41$ \\[5mm]
\end{tabular}
\parbox{0.8\linewidth}{\caption{\label{fig:PNCM_V_syoa}The normalised venue contact matrices (PNCM$_\textbf{V}$)  by age as simulated in \JUNECOX. Note that the data inputs in (i) and (j) stem from a previous survey.}}
\end{figure}
\clearpage
\section{UK Validation}\label{app:UK_Val}
\begin{figure}[h!]
\centering 
\begin{subfigure}[t][][t]{\linewidth}
    \centering 
    \includegraphics[width=\linewidth]{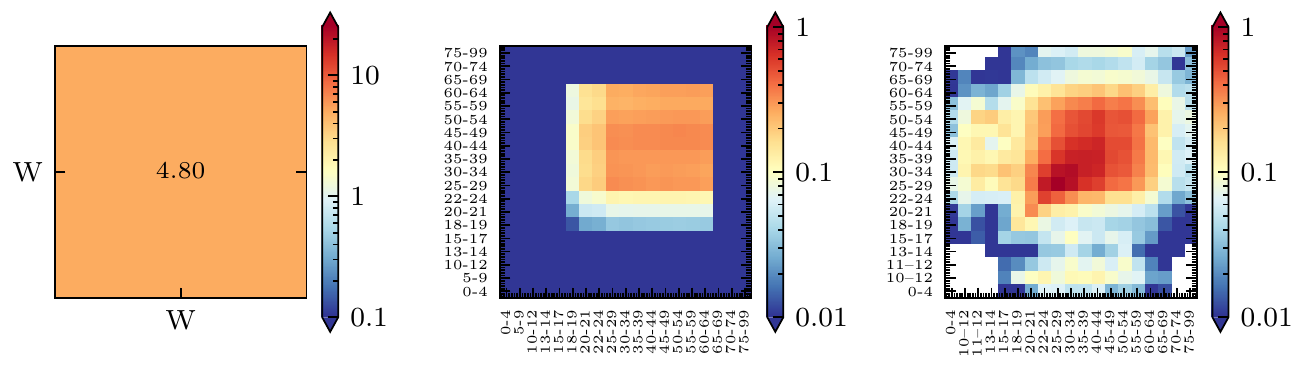}
    \caption{Company}
\end{subfigure}
\begin{subfigure}[t][][t]{\linewidth}
    \centering 
    \includegraphics[width=\linewidth]{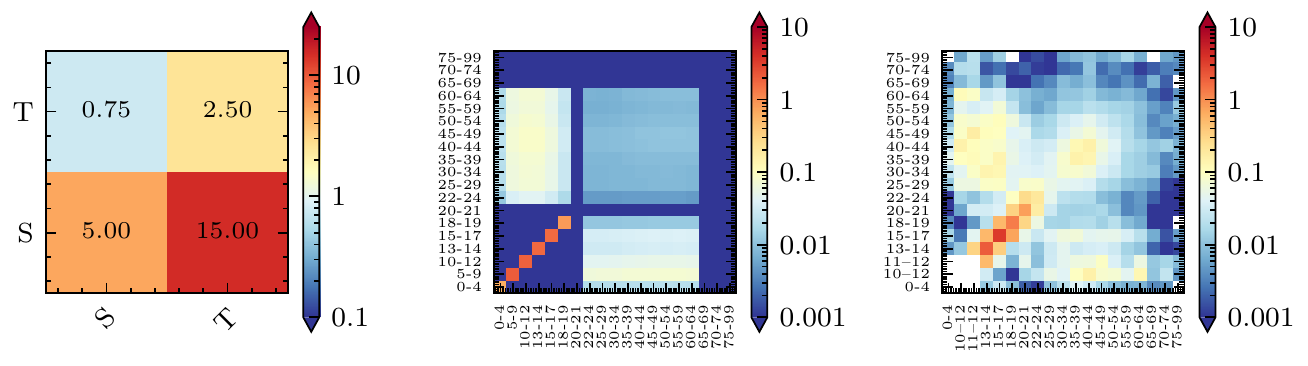}
    \caption{School}
    \end{subfigure}
\parbox{0.8\linewidth}{\caption{\label{fig:CM_UKResult_companySchool} Left: The derived input interaction matrix, UNCM$_\textbf{R}$ for "Companies" and "Schools", where the labels "W" refers to "Workers", "S" students, "T", teachers. 
Center: The simulated age-\-binned PNCM$_\textbf{R}$ matrix with entries $\hat{C}_{ij}$ from \JUNEUK.  
Right: The BBC Pandemic project "all home" contact matrix, $C$, with entries $c_{ij}$.}}
\end{figure} 
\begin{table*}[h!]
\begin{center}
    \begin{tabular}{ | l || c | c | c ||  c |  c | c | } 
    \hline
    \parbox{0.18\textwidth}{$\vphantom{\int\limits_{1}^{12}}$} & \multicolumn{3}{|c||}{Company} & 
    \multicolumn{3}{|c|}{School} \\
    \cline{2-7}
    \parbox{0.15\textwidth}{$\vphantom{\int\limits_{1}^{12}}$} & \parbox{0.1\textwidth}{\hspace*{0.04\textwidth}$Q$} & \parbox{0.1\textwidth}{\hspace*{0.04\textwidth}$I^{2}_{s}$} & \parbox{0.1\textwidth}{\hspace*{0.04\textwidth}$D_{C}$} & 
    \parbox{0.1\textwidth}{\hspace*{0.04\textwidth}$Q$} & \parbox{0.1\textwidth}{\hspace*{0.04\textwidth}$I^{2}_{s}$} & \parbox{0.1\textwidth}{\hspace*{0.04\textwidth}$D_{C}$}\\
     \hline \hline
    $\vphantom{\int\limits^1}$BBC Pandemic & $2.6\times 10^{-2}$ & $0.31$ & & $0.13$ & $0.14$ &  \\[2mm]
    \hline
    $\vphantom{\int\limits^1}$\JUNEUK PNCM$_\textbf{R}$ & $2.6\times 10^{-3}$  & $0.42$ & $0.73$ &  $0.21$  & $5.0 \times 10^{-2}$ & $0.63$\\[2mm]
    \hline
    \end{tabular}
    \\[2mm]
\parbox{0.8\linewidth}{\caption{\label{tab:CompanySchoolUK}Contact matrix statistics calculated for \JUNE-UK and BBC Pandemic project reported for company and school mixing. These statistics are calculated for the UK demography reported by ONS in 2011 \cite{ONS_UK_PopCenus}.}}
\end{center}
\end{table*}

\begin{figure}[h!]
\centering
\includegraphics[width=.9\linewidth]{figures/colourbar_PNCM_V_syoa.pdf}
\begin{tabular}{lllllllll}
\multicolumn{3}{l}{\includegraphics[width=0.3\linewidth]{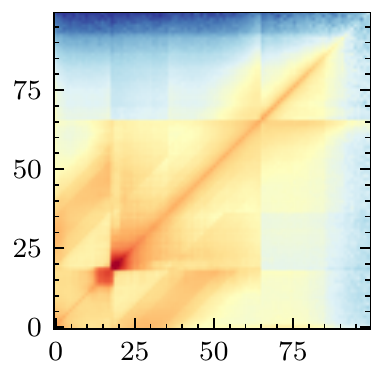}} &
\multicolumn{3}{l}{\includegraphics[width=0.3\linewidth]{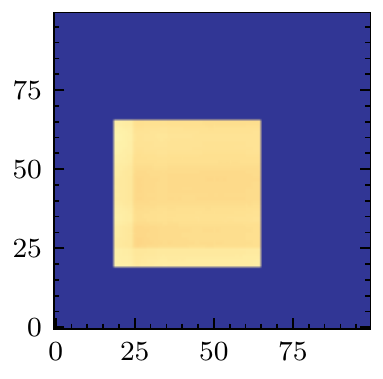}} &
\multicolumn{3}{l}{\includegraphics[width=0.3\linewidth]{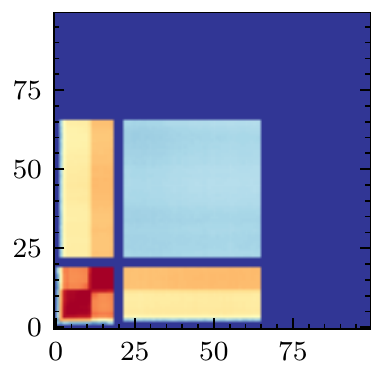}}\\[-1mm]
\hspace*{2mm} & (a) & Household: &
\hspace*{2mm} & (b) & Company:  &
\hspace*{2mm} & (c) & School:\\
&& $Q=4.9\times10^{-2}$ &&& $Q=5.8\times10^{-4}$ &&& $Q=1.3\times10^{-2}$\\
&& $I_{s}^{2}=0.16$ &&& $I_{s}^{2}=0.45$ &&& $I_{s}^{2}=0.050$\\[5mm]
\end{tabular}
\parbox{0.8\linewidth}{\caption{\label{fig:UK_PNCM_V_syoa}The normalised venue contact matrices (PNCM$_\textbf{V}$)  by age as simulated in \JUNEUK.}}
\end{figure}
\clearpage

\section{Survey}\label{app:survey}
The survey between October-November 2020 was conducted by enumerators from the UNHCR Community Based Protection team who regularly conduct surveys within the settlement following standard UNHCR practices~\cite{unhcr_data_collection, unhcr_data_protection}. 
Data was collected from 22 camps in the Kutapalong-Balukhali Expansion Site (part of the Cox's Bazar refugee settlement) consisting of 2 men and 2 women in each of the following categories: $< 18$ years; $\geq 18$ years $< 60$; $\geq 60$ years. In addition 2 persons with disabilities were surveyed to make a total of 308 respondents. Anonymised results, and additional metadata, can be accessed through UNHCR~\cite{microdata_survey}.

The survey was conducted by enumerators randomly sampling households in each camp and visiting them in person. Only one respondent per household was permitted and responses were collected using the Kobo Toolbox~\cite{kobo_toolbox} based on the Open Data Kit~\cite{open_datakit}. The survey was formatted as follows (italicised text is spoken):

\vspace*{5mm}
\textit{This questionnaire has been designed by teams from United Nations Global Pulse and UNHCR and is to inform efforts to better understand how people move around in the camp and interact with others to better understand how COVID-19 might spread in the camp to inform future COVID-19 protection measures.}

\textit{Good day by name is $\textunderscore\textunderscore\textunderscore\textunderscore$ from UNHCR and I am here to conduct a survey. This study is part of a scientific research project from United Nations Global Pulse and UNHCR. In this study, we will ask questions to better understand how people move around in the camp and interact with others. Your decision to complete this study is completely voluntary, and you may decline to answer at any time. Your answers will be completely anonymous. The results of the research may be presented at scientific meetings or published in scientific journals. For any questions or comments please contact: $\textunderscore\textunderscore\textunderscore\textunderscore$. The survey should not take longer than 30 minutes.}

\begin{enumerate}
    \item
    \begin{itemize}
        \item \textbf{If adult:} \textit{Do you declare that you are at least 18 years of age and that you agree to complete this survey voluntarily?}
        \item \textbf{If child:}
        \begin{itemize}
            \item \textbf{To parent or guardian:} \textit{Do you declare that you are at least 18 years of age, that you are the parent or guardian of this child and that you give consent for your child to complete this survey voluntarily?}
            \item \textbf{To child:} \textit{Do you declare that this is your parent or guardian and that you give consent to complete this survey voluntarily?}
        \end{itemize}
    \end{itemize}
    
    \item \textit{Sex:} Female, Male, Other, Do not want to answer
    \item \textit{Location at the time:} $\textunderscore\textunderscore\textunderscore\textunderscore$ (camp)
    \item \textit{Age:} under 18, over 18 but under 60, over 60
    \item \textit{Disability:} Y/N
    \item \textit{Do you have access to a face mask?} Y/N
    
    \item \textit{When the learning centres were open, did you attend any formal education?} Y/N
    \item 
    \begin{itemize}
        \item \textbf{If yes:} 
        \begin{enumerate}
            \item \textit{When you attended formal education, how much time do you spend there?} 30 minutes, 1 hour, 1 hour and 30 minutes, 2 hours, other (please specify)
            \item \textit{When you attended formal education, approximately how many children do you come into contact with (for example, talk to)?}
            \item \textit{When you attended formal education, approximately how many adults do you come into contact with (for example, talk to)?}
        \end{enumerate}
    \end{itemize}
    
    \item \textit{Do you ever go to a food distribution center?} Y/N
    \item 
    \begin{itemize}
        \item \textbf{If yes:} 
        \begin{enumerate}
            \item \textit{When you go to a food distribution center, how much time do you spend there? } 30 minutes, 1 hour, 1 hour and 30 minutes, 2 hours, other (please specify)
            \item \textit{When you go to a food distribution center, approximately how many children do you come into contact with at the center (for example, talk to)?}
            \item \textit{When you go to a food distribution center, approximately how many adults do you come into contact with at the center (for example, talk to)?}
            \item \textit{When you go to the food distribution center, do you wear a mask in the center?}
        \end{enumerate}
    \end{itemize}

    \item \textit{Do you ever go to an e-voucher outlet?} Y/N
    \item 
    \begin{itemize}
        \item \textbf{If yes:} 
        \begin{enumerate}
            \item \textit{When you go to an e-voucher outlet, how much time do you spend there?} 30 minutes, 1 hour, 1 hour and 30 minutes, 2 hours, other (please specify)
            \item \textit{When you go to an e-voucher outlet, approximately how many children do you come into contact with at the outlet (for example, talk to)?}
            \item \textit{When you go to an e-voucher outlet, approximately how many adults do you come into contact with at the outlet (for example, talk to)?}
            \item \textit{When you go to an e-voucher outlet, do you wear a mask in the outlet?}
        \end{enumerate}
    \end{itemize}
    
    \item \textit{Do you ever go to a community center?} Y/N
    \item 
    \begin{itemize}
        \item \textbf{If yes:} 
        \begin{enumerate}
            \item \textit{When you go to a community center, how much time do you spend there?} 30 minutes, 1 hour, 1 hour and 30 minutes, 2 hours, other (please specify)
            \item \textit{When you go to a community center, approximately how many children do you come into contact with at the center (for example, talk to)?}
            \item \textit{When you go to a community center, approximately how many adults do you come into contact with at the center (for example, talk to)?}
            \item \textit{When you go to a community center, do you wear a mask in the center?}
        \end{enumerate}
    \end{itemize}
    
    \item \textit{Do you ever go to a religious meeting?} Y/N
    \item 
    \begin{itemize}
        \item \textbf{If yes:} 
        \begin{enumerate}
            \item \textit{When you go to a religious meeting, how much time do you spend there?} 30 minutes, 1 hour, 1 hour and 30 minutes, 2 hours, other (please specify)
            \item \textit{When you go to a religious meeting, approximately how many children do you come into contact with at the meeting (for example, talk to)?}
            \item \textit{When you go to a religious meeting, approximately how many adults do you come into contact with at the meeting (for example, talk to)?}
            \item \textit{When you go to a religious meeting, do you wear a mask in the meeting?}
        \end{enumerate}
    \end{itemize}
    
    \item 
    \begin{enumerate}
        \item \textit{When you go to a water pump or latrine, how much time do you spend there?} 30 minutes, 1 hour, 1 hour and 30 minutes, 2 hours, other (please specify)
        \item \textit{When you go to a water pump or latrine, approximately how many children do you come into contact with (for example, talk to)?}
        \item \textit{When you go to a water pump or latrine, approximately how many adults do you come into contact with (for example, talk to)?}
        \item \textit{When you go to a hand pump or latrine, do you wear a mask?}
    \end{enumerate}    
\end{enumerate}

\section{Questions for the CBP team}\label{app:Questions}

To supplement our analysis, a series of informal interviews were conducted with members of the Cox's Bazar refugee settlement UNHCR Community Based Protection (CBP) team. In each of these interviews a set of general enquires into the behaviour and attendance rates were asked of members of the protection team which worked closely with those venue types.

\vspace*{5mm}
\textit{This questionnaire has been designed by teams from United Nations Global Pulse and UNHCR and is to inform efforts to better understand how people engage with each venue in the camp and the demography of the venues.}

\textit{For the following venues: Community centres, Female friendly spaces, Food distribution centres, E-voucher outlets, Non-food distribution centres - including LPG and blanket centres - Religious centres, and Learning centres. Where you are able and suitably informed please could you answer the following questions;}

\begin{enumerate}
    \item
    \begin{itemize}
        \item \textit{Can you describe what a day looks like at} \textbf{venue} \textit{?}
        \begin{itemize}
            \item \textit{How many people do you expect at minimum and peak times?}
            \item \textit{How do these days and numbers of people vary by day, week, month/season?}
            \item \textit{Why do you think there are these variations?}
        \end{itemize}
    \end{itemize}
    \item
    \begin{itemize}
        \item \textit{What is the makeup of multigenerational households - are there generally three generations or more?}
        \begin{itemize}
            \item \textit{Do these households include extended family?}
            \item \textit{Is this a cultural issue or a space constraint?}
        \end{itemize}
    \end{itemize}
    \item
    \begin{itemize}
        \item \textit{What age do children typically move through the camp independently?}
        \begin{itemize}
            \item \textit{Move out from parents shelter?}
            \item \textit{Go to} \textbf{venues} \textit{on their own? (e.g collect items from the distribution centres for their shelter)}
            \item \textit{How many hours do they spend moving around in the camp independently?}
            \item \textit{Who do they mostly have contact with when they move around? (e.g. more children, teachers at school, other adults? all)}
        \end{itemize}
    \end{itemize}
    \item
    \begin{itemize}
        \item \textit{What time do} \textbf{venues} \textit{close?}
    \end{itemize}
\end{enumerate}
\clearpage
\begin{figure}[h!]
\centering
\begin{subfigure}[t][][t]{.45\linewidth}
    \centering
    \includegraphics[width=\linewidth]{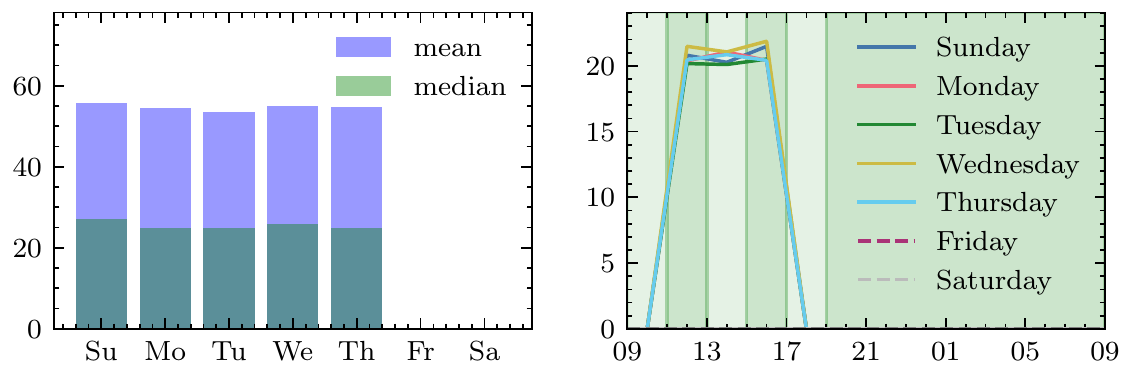}
    \caption{Communal spaces}
\end{subfigure}
\begin{subfigure}[t][][t]{.45\linewidth}
    \centering
    \includegraphics[width=\linewidth]{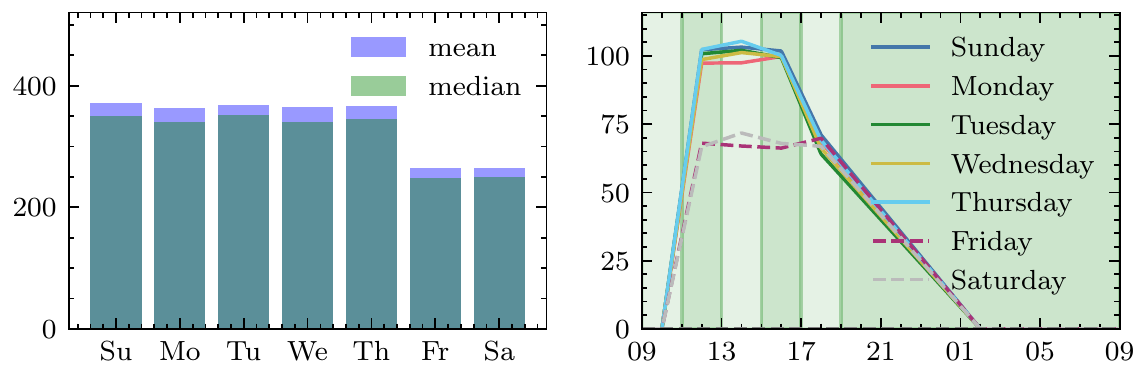}
    \caption{Distribution centres}
\end{subfigure}
\begin{subfigure}[t][][t]{.45\linewidth}
    \centering
    \includegraphics[width=\linewidth]{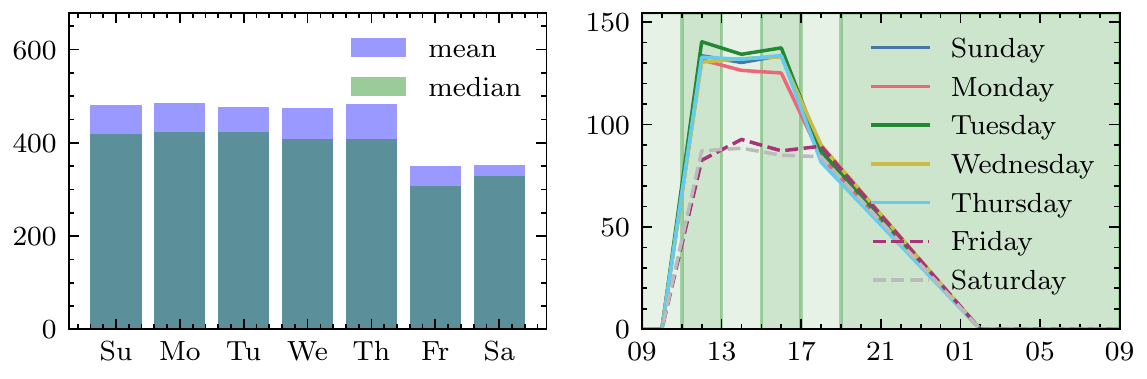}
    \caption{e-voucher outlets}
\end{subfigure}
\begin{subfigure}[t][][t]{.45\linewidth}
    \centering
    \includegraphics[width=\linewidth]{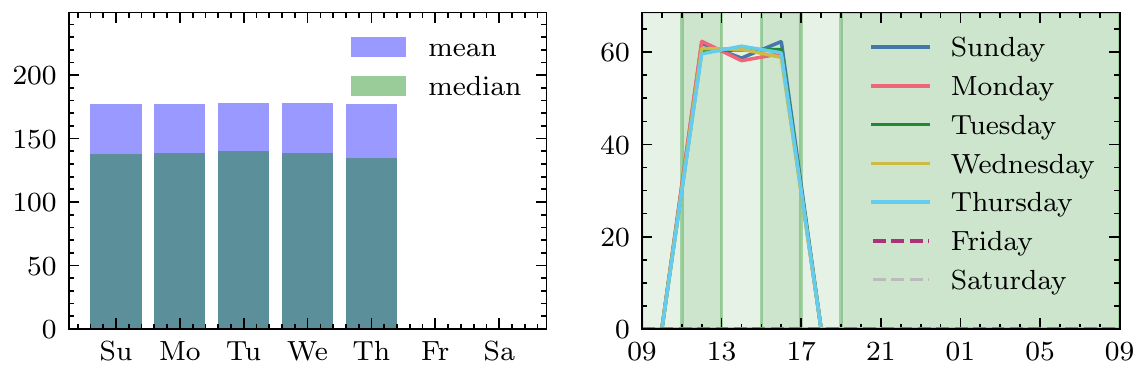}
    \caption{Female friendly spaces}
\end{subfigure}
\begin{subfigure}[t][][t]{.45\linewidth}
    \centering
    \includegraphics[width=\linewidth]{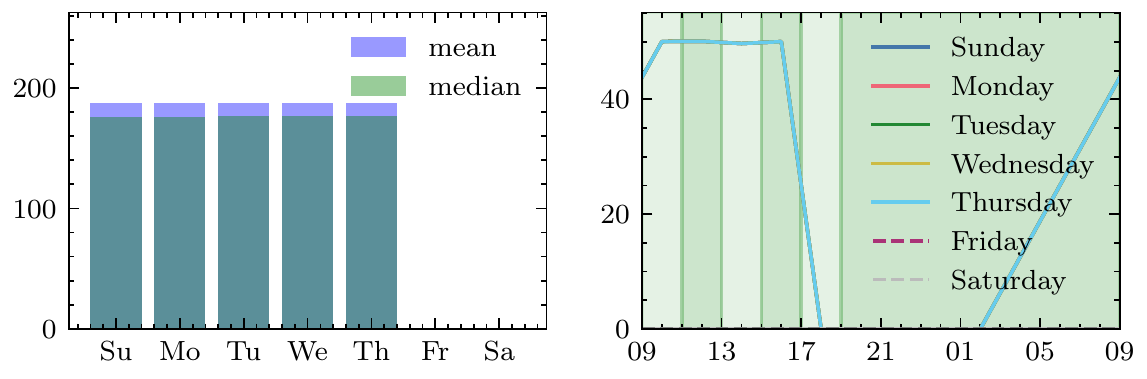}
    \caption{Learning centres}
\end{subfigure}
\begin{subfigure}[t][][t]{.45\linewidth}
    \centering
    \includegraphics[width=\linewidth]{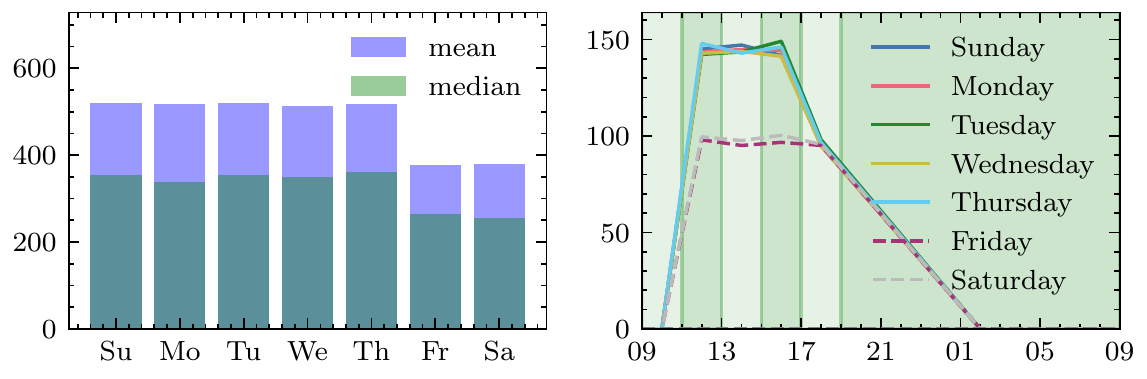}
    \caption{Non-food distribution centres}
\end{subfigure}
\begin{subfigure}[t][][t]{.45\linewidth}
    \centering
    \includegraphics[width=\linewidth]{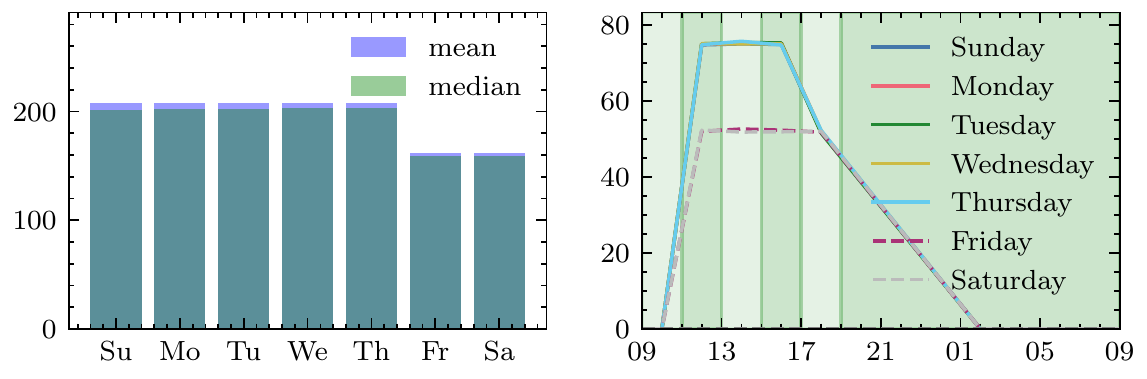}
    \caption{Pumps and Latrines}
\end{subfigure}
\begin{subfigure}[t][][t]{.45\linewidth}
    \centering
    \includegraphics[width=\linewidth]{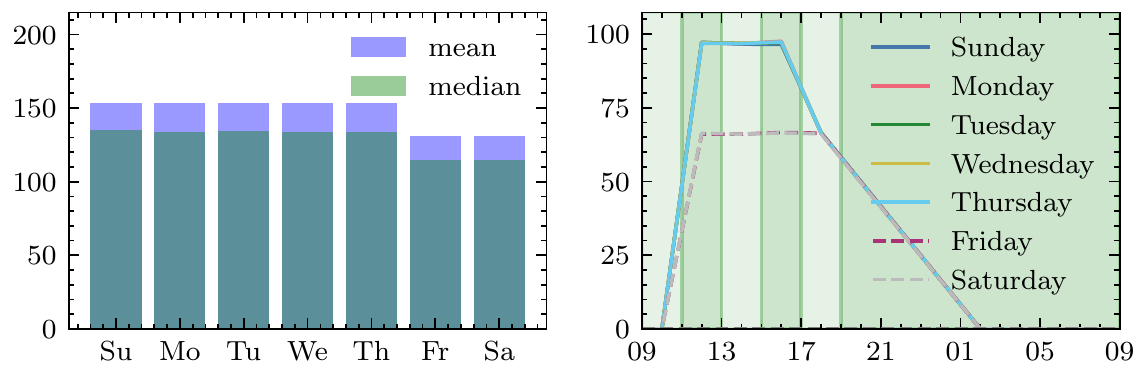}
    \caption{Religious centres}
\end{subfigure}
\parbox{0.8\linewidth}{\caption{\label{fig:Populations} The unique person attendance rates per day (left) and by time of day (right) for the virtual venue. The green shading represents the discrete timestep bins of the simulation.}}
\end{figure} 

The attendance probabilities (see Figure \ref{fig:AgeProbabilities}) were tuned to achieve the desired attendance rates. These rates were chosen such that they represent an "average" day of any particular day of the week in the camp ignoring any changes of behaviour from religious or national events or annual variations in climate and weather. 

\section{Demographic properties}\label{app:DemographicProperties}
\begin{figure}[h!]
\centering 
\includegraphics{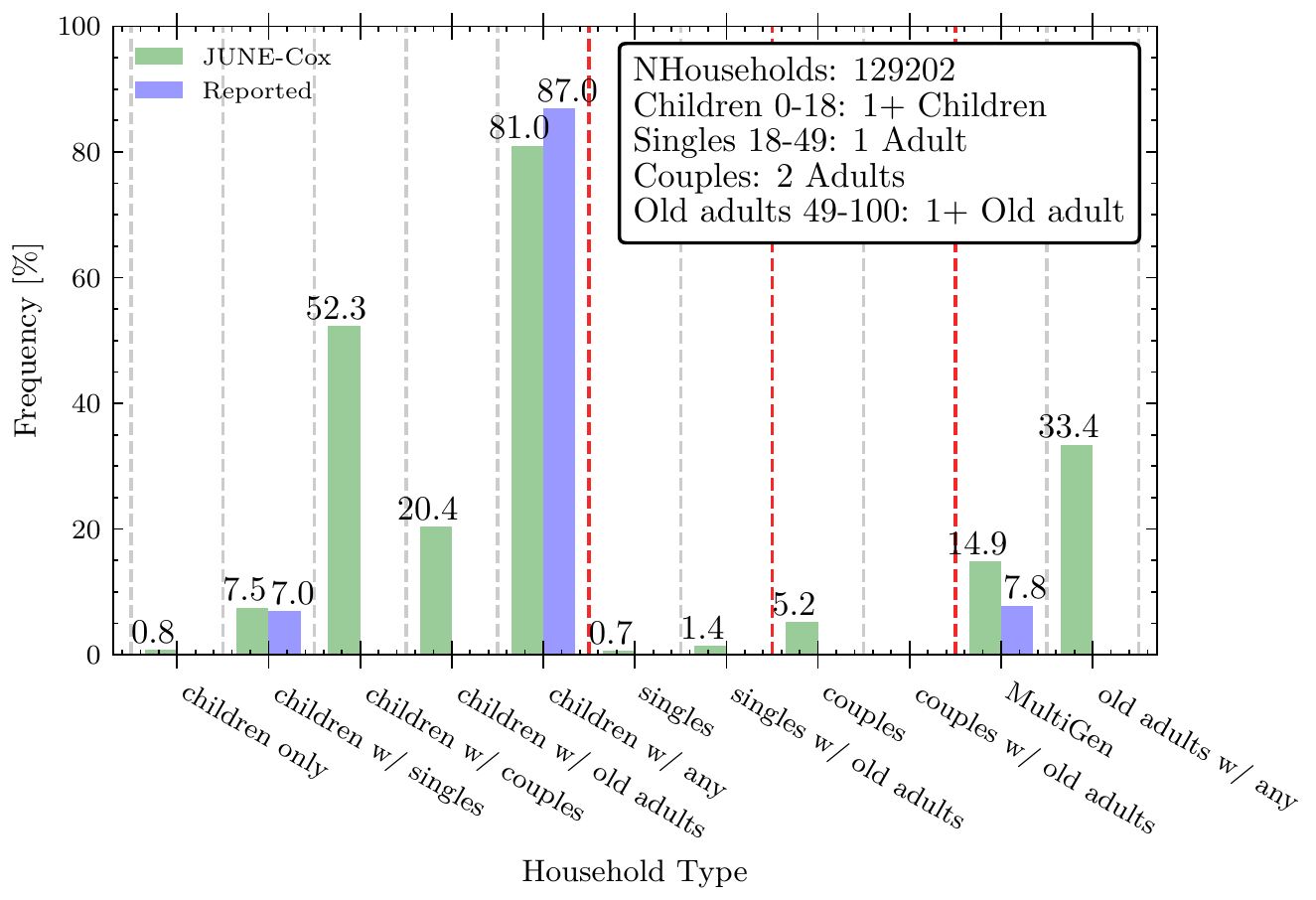}
\parbox{0.8\linewidth}{\caption{\label{fig:HouseTypes} Figure of proportion of household types. Note that not all of these groups are mutually exclusive. Green represents the reconstruction in \JUNE and Blue the reported data (if available). Those groups where data was unavailable are reported in the figure for completeness. }}
\end{figure}

\begin{figure}[h!]
\centering 
\includegraphics{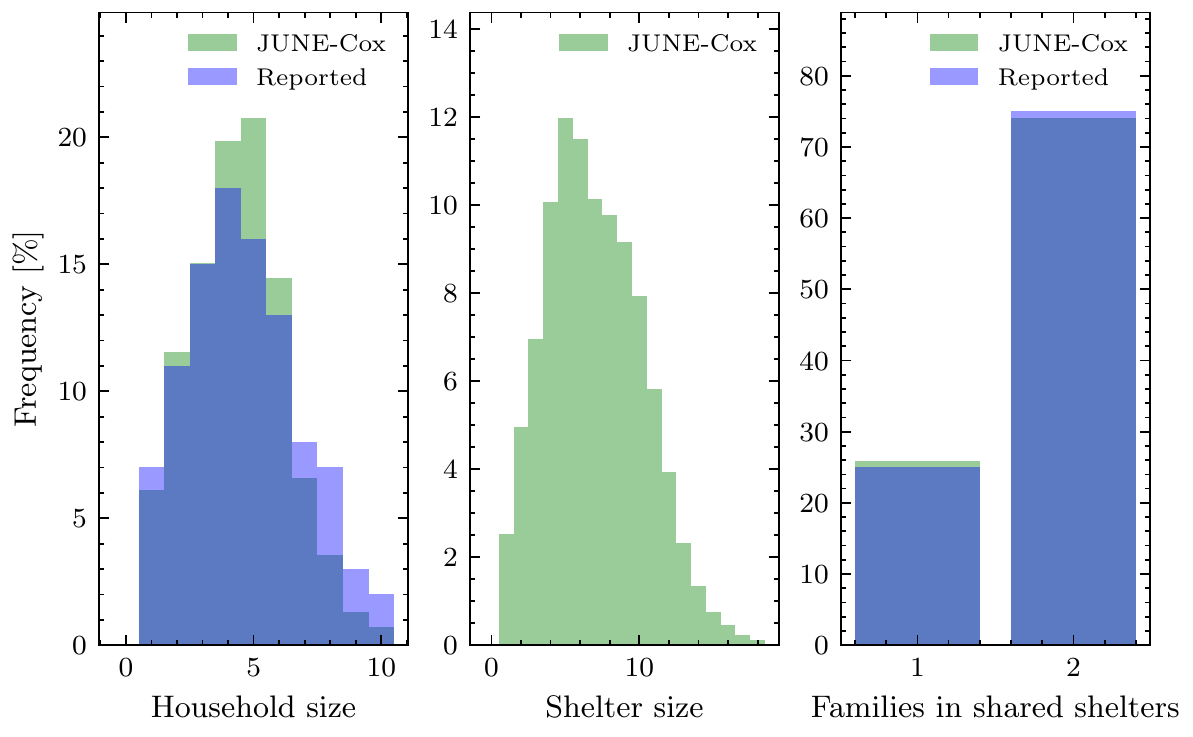}
\parbox{0.8\linewidth}{\caption{\label{fig:HouseholdSizes} Figure of key shelter properties. Left: distribution of household family sizes. Middle: distribution of shelter sizes. Right: Proportion of one and two household shelters. Green represents the reconstruction in \JUNE and Blue the reported data (if available). }}
\end{figure}

\begin{figure}[h!]
\centering 
\includegraphics{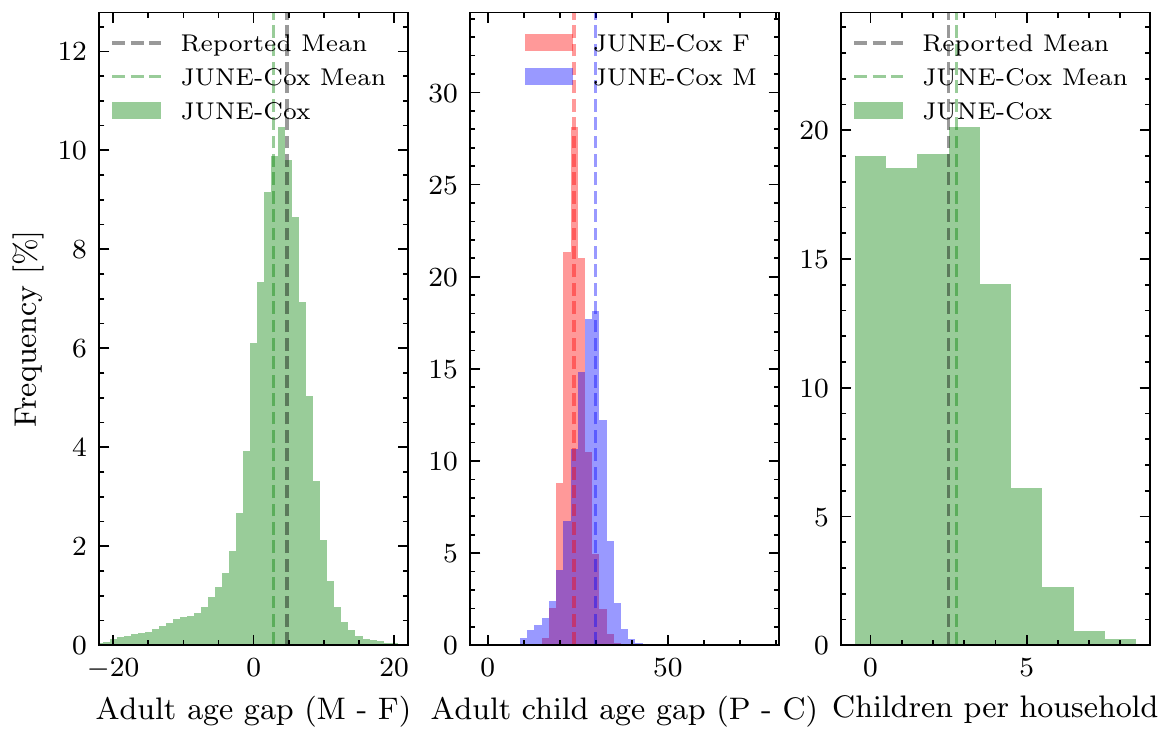}
\parbox{0.8\linewidth}{\caption{\label{fig:DemographicProperties} Figure of microscopic household properties. Left: distribution of household male female age gap in houses containing only two adults in the range [24-49]. Middle: distribution adult child age gaps in households containing one adult [24-49] and the eldest child [0-18]. Right: distribution of number of children by household. Green represents the reconstruction in \JUNECOX and black dashed lines the reported mean data (if available). }}
\end{figure}

Households are constructed stochastically by clustering individuals into households according to their age, sex and the following reported properties of the camp in order to create realistic demographic household structures:
\begin{itemize}
    \item Macroscopic properties:
    \begin{enumerate}
        \item The distribution of household sizes in the camp (known at the region level);
        \item Population demographics;
        \item The proportion of one, two and multi-generational households.
    \end{enumerate}
    \item Microscopic properties:
    \begin{enumerate}
    \item The likelihood of single parent;
    \item The mean spousal age gap;
    \item The mean age of mother at birth of first child.
    \end{enumerate}
\end{itemize}
These properties are all known at the super-area level unless specified otherwise.
The resulting household demographic structures can be seen in Figure \ref{fig:HouseTypes} and shelter sizes in Figure \ref{fig:HouseholdSizes}. The age brackets for each demographic are inferred from survey and data from the settlement. Children [0 - 18], 18 is the age at which marriage is legal for women (21 for men), Adults [18 - 49] (49 being chosen to provide a realistic age gap for potential grandparents, twice the average mother-child age gap, 22.43 years plus the average spousal age gap, 4.73 years). [49 - 100] for old adults, the remaining ages in the camp. \JUNECOX has an over clustering of children with single parent housing, this due to any remaining children being randomly clustered into households with adults after the children with couples houses are constructed. The microscopic properties of the clustered households are summarised between Figures \ref{fig:HouseTypes} and \ref{fig:DemographicProperties}.

\clearpage
\section{Algorithm for the virtual survey}
\label{app:algo_virtual_survey}
\RestyleAlgo{ruled}
\begin{algorithm}
    \SetKwInput{KwInput}{Input}                
    \SetKwInput{KwOutput}{Output}              
    \DontPrintSemicolon
    \parbox{0.8\linewidth}{
        \caption{
            \label{alg:CountContacts}
            The virtual survey. 
            Loop over all venues and people and simulate $P_{\textrm{contacts}}$ between $i$ and $j$ subgroups from survey. 
            The contacts can then be clustered into arbitrary subgroups $k, l$. 
            We allow for multiple contacts between the same people at venue $L$.
        }
    }
    \KwData{$\hat{t}_{ij}^{L}=[0]_{kl}$} 
    \For {$L \in$ {\rm Venues}}{ 
        \For{$P_{x} \in$ {\rm People} @ $L$, $P^L$}
            {
            $i = \mathrm{subgroup}(P_x)$\\
            $T^L = T^L + \Delta T$\\
            $\hat{\eta}^L_{i} = \hat{\eta}^L_{i} + 1$
            }
        \For{$j \in L_{\mathrm{subgroups}}$}
            { 
            Generate $\Tilde{\gamma}^L_{ij}$
            \eIf 
                {$\Tilde{\gamma}^L_{ij} = 0$}
                { continue }
                {Generate randomly a list of $P_{\mathrm{contacts}}$ of $\Tilde{\gamma}^L_{ij}$ people at $L$ in subgroup $j$ not including $P_x$}
            }
        \For {$P_c \in P_{\mathrm{contacts}}$}
            {
            $k = \mathrm{subgroup}(P_x)$\\
            $l = \mathrm{subgroup}(P_c)$\\
            $\hat{t}^{L}_{kl} = \hat{t}^{L}_{kl} + 1$
            }
    }
    $\hat{t}^{L}_{kl} = \hat{t}^{L}_{kl} / T^{L} $
\end{algorithm}

\end{document}


\flushbottom
\maketitle

\section{Taxonomies}\label{appendix:taxonomies}


\subsection{Taxonomy for benchmarking ASRs}
\label{appendix:taxonomies_benchmarking_asr}

{Words related to the COVID-19 pandemic:}
\begin{greybox}
\texttt{symptom OR inoculat OR transmi OR vaccin* OR variant OR virus OR doses OR delta OR oxford OR "johnson and" OR "first wave" OR "second wave" OR immun* OR infect* OR contract* OR quarantine OR pandemic OR preventative OR "side effect"}
\end{greybox}

\bigskip

\noindent{Variations of the term COVID-19:}
\begin{greybox}
\texttt{"covert nineteen" OR "covered nineteen" OR "cove nineteen" OR "clover nineteen" OR "covet nineteen OR "coffee nineteen" OR coronated}
\end{greybox}

\bigskip

\subsection{Taxonomy for Case Study 2}

\begin{greybox}
\texttt{(("herbal treatment" OR "herbal medicine" OR “remedy” OR “remedies” OR "herbal drug" OR "covidex" OR "codivex" OR "covilyce" OR "covilyce-1" OR "covylice" OR “codivex” OR “covylice”) AND (coronavirus OR covid OR "sars cov 2" OR "corona virus" OR "covid 19" OR "cov 19" OR "covid nineteen" OR "corona" OR "covid-19"))}
\end{greybox}

\bigskip

\subsection{Taxonomy for Case Study 3}

{Omicron in general:}
\begin{greybox}
\texttt{omicron OR acron OR unicron OR omiclon OR omikron}
\end{greybox}

\noindent{Skepticism for vaccine effectiveness:}
\begin{greybox}
\texttt{(omicron OR acron OR unicron OR omiclon OR omikron) AND (vaccine OR vaccines OR vaccina* OR jabs OR "jab" OR antivax OR antivax* OR "no-vax" OR "no-vax*" OR immunisation OR "vaccination" OR vaccin OR "vaccins" OR "anti-vaccin" OR "refus vaccin*" OR "inocul*") AND ("not effective" OR effectiveness)}
\end{greybox}

\section{Radio Stations}

\input{tables/appendix_radio_stations_1_3}
\input{tables/appendix_radio_stations_2}

\bibliography{references}